%% file: main.tex
\documentclass[sigconf]{acmart}

\copyrightyear{2022} 
\acmYear{2022} 
\setcopyright{rightsretained} 
\acmConference[WWW '22]{Proceedings of the ACM Web Conference 2022}{April 25--29, 2022}{Virtual Event, Lyon, France} 
\acmBooktitle{Proceedings of the ACM Web Conference 2022 (WWW '22), April 25--29, 2022, Virtual Event, Lyon, France}
\acmDOI{10.1145/3485447.3512226} 
\acmISBN{978-1-4503-9096-5/22/04}

\AtBeginDocument{%
  \providecommand\BibTeX{{%
    \normalfont B\kern-0.5em{\scshape i\kern-0.25em b}\kern-0.8em\TeX}}}
\usepackage{nicefrac}
\usepackage{siunitx}
\usepackage{array,framed}
\usepackage{booktabs}
\usepackage{
  color,
  float,
  epsfig,
  wrapfig,
  graphics,
  graphicx,
  subcaption
}
\usepackage{textcomp}

\usepackage{amssymb}
\usepackage{setspace}
\usepackage{latexsym,fancyhdr,url}
\usepackage{enumerate}
\usepackage{algorithm2e}
\usepackage{algpseudocode}
\usepackage{graphics}
\usepackage{xparse} % argument parsing -- \edist
\usepackage{xspace}
\usepackage{multirow}
\usepackage{csvsimple}
\usepackage{balance}
\usepackage{multirow}
\usepackage{url}
\usepackage{graphicx}
\usepackage{
  tikz,
  pgfplots,
  pgfplotstable
}
\usepackage{hyperref}

\usetikzlibrary{
  shapes.geometric,
  arrows,
  external,
  pgfplots.groupplots,
  matrix
}

\pgfplotsset{compat=1.9}
\usepackage{mathtools,}

\DeclareMathAlphabet{\mathcal}{OMS}{cmsy}{m}{n}
\DeclareGraphicsExtensions{%
    .png,.PNG,%
    .pdf,.PDF,%
    .jpg,.mps,.jpeg,.jbig2,.jb2,.JPG,.JPEG,.JBIG2,.JB2}

\input{defs}
\begin{document}
%\fontfamily{lmr}\selectfont
% \def\thetitle{A Practical Way to Generate Strong Keys from Noisy Data}
% \fancyhead{}
\def\thetitle{TTAGN: Temporal Transaction Aggregation Graph Network for Ethereum Phishing Scams Detection}

\title{\thetitle}

% \begin{comment}
% \author{Sijia Li\textsuperscript{1,2}, Gaopeng Gou\textsuperscript{1,2}, Chang Liu$^{1,2\ast}$, Chengshang Hou\textsuperscript{1,2}, Zhenzhen Li\textsuperscript{1,2}, Gang Xiong\textsuperscript{1,2}}

\author{Sijia Li$^{1,2}$, Gaopeng Gou$^{1,2}$, Chang Liu$^{1,2\ast}$, Chengshang Hou$^{1,2}$, Zhenzhen Li$^{1,2}$, Gang Xiong$^{1,2}$}

\affiliation[obeypunctuation=true]{
\textsuperscript{1}
{\institution{Institute of Information Engineering, Chinese Academy of Sciences}\country{, Beijing, China}}
}
\affiliation[obeypunctuation=true]{
\textsuperscript{2}
{\institution{School of Cyber Security, University of Chinese Academy of Sciences}\country{, Beijing, China}}
}
% \affiliation{
%   \institution{\textsuperscript{1}Institute of Information Engineering, Chinese Academy of Sciences, Beijing, China}}
% \affiliation{
%   \institution{\textsuperscript{2}School of Cyber Security, University of Chinese Academy of Sciences, Beijing, China}}
\thanks{\textsuperscript{$\ast$}Chang Liu is the corresponding author.}

\email{{lisijia, gougaopeng, liuchang, houchengshang, lizhenzhen, xionggang}@iie.ac.cn}

\ccsdesc[500]{Applied computing~Digital cash}
\ccsdesc[500]{Security and privacy~Phishing}

\keywords{Blockchain, Ethereum, Phishing scams detection, Network representation learning}

\date{}

% \settopmatter{printfolios=true}
\renewcommand{\shortauthors}{Sijia Li et al.}
\settopmatter{printacmref=true}
\input{abstract}
\maketitle

% Section I
\input{intro}    % basic introduction
\input{related_work}

\input{problem_definition}
\input{methodology}

\input{experiments}

\input{conclusion}

\input{acknowledgments}

\bibliographystyle{ACM-Reference-Format}
\bibliography{bib}

% % --- Appendix ---%
% \appendix
% \input{appendix}

\end{document}

%% file: defs.tex
\usepackage{xparse}
\newcommand{\bnm}{\begin{newmath}}
\newcommand{\enm}{\end{newmath}}

\newcommand{\bea}{\begin{eqnarray*}}%
\newcommand{\eea}{\end{eqnarray*}}%

\newcommand{\bne}{\begin{newequation}}
\newcommand{\ene}{\end{newequation}}

\newcommand{\bal}{\begin{newalign}}
\newcommand{\eal}{\end{newalign}}

\newenvironment{newalign}{\begin{align}%
\setlength{\abovedisplayskip}{4pt}%
\setlength{\belowdisplayskip}{4pt}%
\setlength{\abovedisplayshortskip}{6pt}%
\setlength{\belowdisplayshortskip}{6pt} }{\end{align}}

\newenvironment{newmath}{\begin{displaymath}%
\setlength{\abovedisplayskip}{4pt}%
\setlength{\belowdisplayskip}{4pt}%
\setlength{\abovedisplayshortskip}{6pt}%
\setlength{\belowdisplayshortskip}{6pt} }{\end{displaymath}}

\newenvironment{newequation}{\begin{equation}%
\setlength{\abovedisplayskip}{4pt}%
\setlength{\belowdisplayskip}{4pt}%
\setlength{\abovedisplayshortskip}{6pt}%
\setlength{\belowdisplayshortskip}{6pt} }{\end{equation}}

\newcounter{ctr}

\newcounter{mytable}
\def\mytable{\begin{centering}\refstepcounter{mytable}}
\def\endmytable{\end{centering}}

\newcounter{myfig}
\def\myfig{\begin{centering}\refstepcounter{myfig}}
\def\endmyfig{\end{centering}}

\newlength{\saveparindent}
\newlength{\saveparskip}
\newcommand{\E}{{\rm I\kern-.3em E}}

\renewcommand{\eqref}[1]{\mbox{Equation~(\ref{#1})}}

\def \part {part}

\renewcommand{\paragraph}[1]{\vspace*{6pt}\noindent\textbf{#1}\;}

\def \blackslug{\hbox{\hskip 1pt \vrule width 4pt height 8pt
    depth 1.5pt \hskip 1pt}}
\def \qed{\quad\blackslug\lower 8.5pt\null\par}
\newcounter{mynote}[section]

\newcommand\ignore[1]{}

\newcounter{rcnote}[section]

\newcounter{mrnote}[section]

\newcounter{fknote}[section]

\newcounter{anote}[section]

\DeclareMathSymbol{\mlq}{\mathord}{operators}{``}
\DeclareMathSymbol{\mrq}{\mathord}{operators}{`'}

\newcommand{\rhf}[2]{R_{f, \gamma}}

\DeclareDocumentCommand{\edist}{o o}{
  \ensuremath{
    \IfNoValueTF{#1}{{d}}{{\sf d}(#1,#2)}
  }
}

\newcommand{\olrk}[1]{\ifx\nursymbol#1\else\!\!\mskip4.5mu plus 0.5mu\left(\mskip0.5mu plus0.5mu #1\mskip1.5mu plus0.5mu \right)\fi}

\NewDocumentCommand{\indseq}{ O{1} O{r} }{{#1}\ldots {#2}}

%% file: abstract.tex
\begin{abstract}
% In recent years, phishing scams have become the largest type of crime involved in Ethereum, the second largest blockchain platform. 
In recent years, phishing scams have become the most serious type of crime involved in Ethereum, the second-largest blockchain platform.
%   Because of the diversity of Ethereum phishing scams, the method of phishing detection of fixed webpages or emails in traditional scenarios is no longer applicable. Ethereum transaction data can be naturally abstracted into a complex graph of the real world. 
% The existing phishing detection technology on Ethereum mostly uses traditional machine learning or graph representation learning to learn information about the transaction network. However, these methods only use the last transaction or completely ignore the transaction records between addresses. They use manually extracted statistical features as the attribute representation of the node which weakly for phishing address detection on Ethereum.
The existing phishing scams detection technology on Ethereum mostly uses traditional machine learning or network representation learning to mine the key information from the transaction network to identify phishing addresses. However, these methods adopt the last transaction record or even completely ignore these records, and only manual-designed features are taken for the node representation. 
%   However, the anonymity of the blockchain and the extremely imbalance distribution of label greatly reduce the learning efficiency. 
% To tackle these challenges, we propose Temporal Transaction Aggregation Graph Network (TTAGN) to enhance the performance of phishing scam detection on Ethereum. In our model, we adopt temporal transaction records and model them to enrich the representation of nodes. Moreover, we import and design edge2node to automatically aggregate edge embedding to nodes, which enhances the detection of phishing nodes on Ethereum.
In this paper, we propose a Temporal Transaction Aggregation Graph Network (TTAGN) to enhance phishing scams detection performance on Ethereum. Specifically, in the temporal edges representation module, we model the temporal relationship of historical transaction records between nodes to construct the edge representation of the Ethereum transaction network. Moreover, the edge representations around the node are aggregated to fuse topological interactive relationships into its representation, also named as trading features, in the edge2node module. We further combine trading features with common statistical and structural features obtained by graph neural networks to identify phishing addresses.
%   we make full use of the known massive transaction timing information, convert all transaction information between nodes into temporal sequences and use lstm to model, effectively learning the representation of edges. Next, we use edge embedding to learn the node representation and introduce an attention mechanism to effectively solve the problem of uneven label distribution to a certain extent. Finally, we use the graph autoencoder to further perceive the phishing address structural features. 
% We collected a real-world Ethereum phishing scam dataset, experiment results demonstrate that the TTAGN achieves excellent performances (92.8\% AUC, 85.9\% Recall, 77.7\% Pre and 81.6\% F1-score) and outperforms the state-of-the-art methods. The ablation experiment also verified the effectiveness of the module we designed.
Evaluated on real-world Ethereum phishing scams datasets, our TTAGN  (92.8\% AUC, and 81.6\% F1-score) outperforms the state-of-the-art methods, and the effectiveness of temporal edges representation and edge2node module is also demonstrated.
% label distribution of nodes is heavily skewed
\end{abstract}

%% file: intro.tex
\section{Introduction}
\label{sec:intro}

Ethereum\cite{wood2014ethereum} is one of the most popular and scalable blockchain platforms with 14.8 transactions per second and 700,000 daily active addresses on it\cite{2020survey}.
However, along with its high-speed development, Ethereum has also become a hotbed of various cybercrimes\cite{hotbed}. 
% More than 10\% of Initial coin offering (ICO) released on Ethereum have been reported to be suffer from a variety of scams, including phishing, Ponzi schemes, etc.\cite{russon2017ethereum}. 
Phishing, as a typical scam, has received a great deal of attention due to its high visibility and lots of potential victims.
% According to a report of Chainalysis\footnote{https://www.chainalysis.com/}, a provider of investigation and risk management software for virtual currencies, in the whole year of 2020 and the first half of 2021, phishing scams are the most deceptive form of fraud.  
Based on 2021 statistics of phishing scams from Chainalysis, victims lost \$645,000 within the first week of the phishing campaign, and the attacker’s illegal profits exceeded \$3,000,000 in just one month\cite{chainalysis}. Phishing scams cause great economic losses and have become a major threat to the trading security of Ethereum\cite{2020survey}. Therefore, identifying phishing scams on Ethereum becomes a crucial research topic and attracts widespread attention\cite{survey2019blockchain,wang2021ethereumsurvey}.
% There are still a considerable number of people entering the crypto market, and the report also concluded that , suggesting that the research on phishing scams remains significant\cite{chainalysis}. In general, the phishing scam is an urgent problem, causing great economic losses and fighting with it is a long-term task.
% \footnote{https://www.chainalysis.com/}
\begin{figure}[htbp]
\centering
\includegraphics[width = 0.95\linewidth]{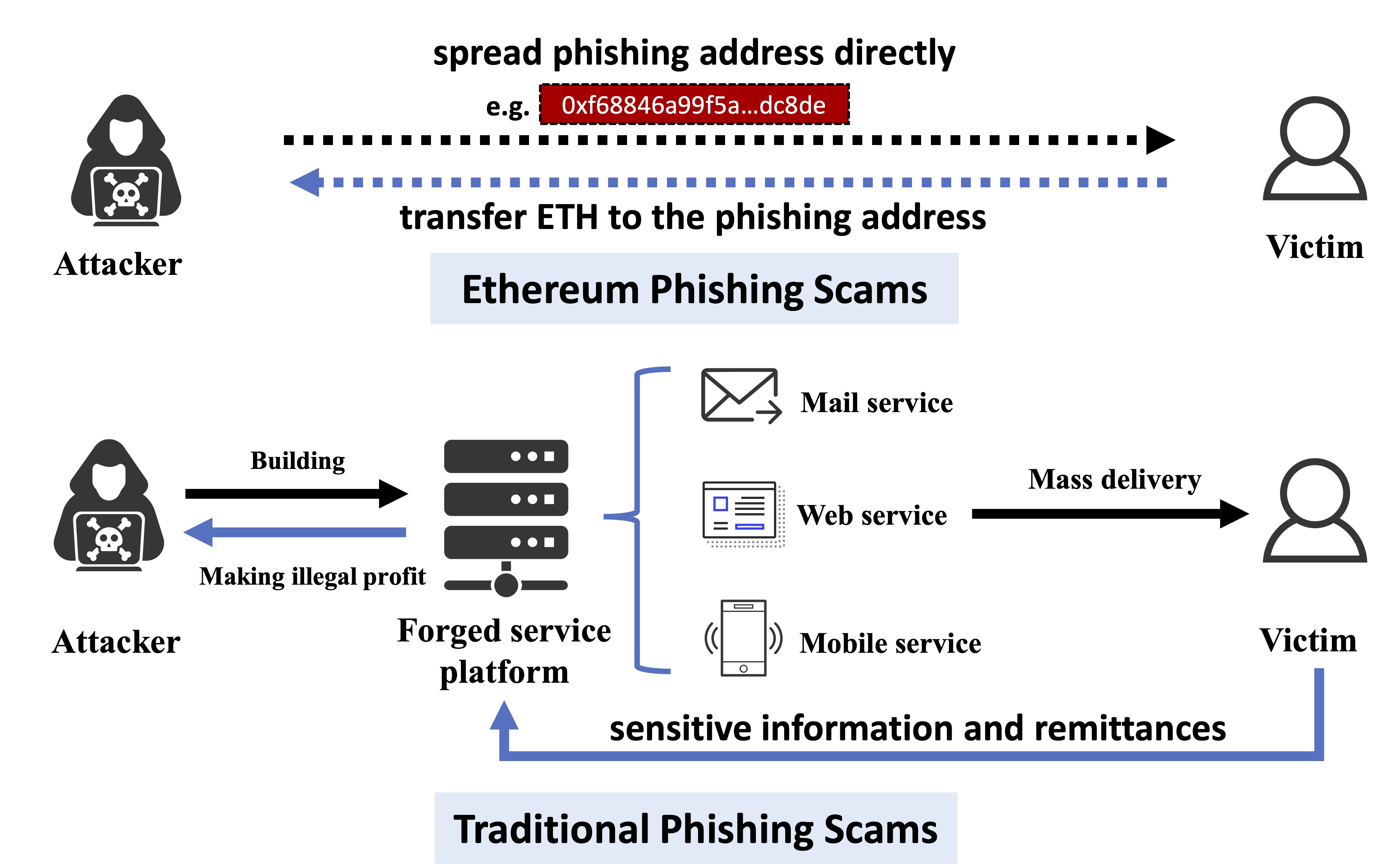}
\caption{The difference between traditional phishing scams and Ethereum phishing scams. }
% Traditional phishing attackers usually build related service platforms to collect sensitive user information and accept remittance, Ethereum phishing scam takes high-reward propaganda to induce remittances, these methods are more flexible and do not have a fixed pattern
\label{difference}
\end{figure}

Traditional phishing scams detection methods cannot be well adapted to the Ethereum scenario. As Figure \ref{difference} shows,
traditional phishing scams rely on building forged platforms (websites or software) to collect sensitive information or receive remittances from victims, so traditional methods focus on mining forged platform patterns, such as CSS styles\cite{haruta2017ccs}, website URLs\cite{sahingoz2019url}, etc. However, in Ethereum, phishing organizations take high-reward propaganda to induce remittances\cite{chainalysis}, they can swindle money directly without forged platforms by spreading phishing addresses to victims in any way such as emails, chat groups, etc. Since there is no fixed pattern for phishing scams on Ethereum, traditional detection methods are ineffective.
The current methods of phishing scam detection on Ethereum are to learn the representation of phishing nodes through the transaction network and classify nodes\cite{lin2020modeling,alqassem2018anti}, in which nodes represent Ethereum transaction addresses and edges represent transactions between addresses. 
The main detection methods can be roughly divided into two types. One is to combine traditional machine learning and manual-designed features (i.e. structural and statistical features) of nodes for phishing detection~\cite{chen2020phishing}. However, these methods mainly rely on professional knowledge to extract manual-designed features 
(e.g., node’s in-degree, total transaction amount, transaction time interval, etc.)
, which are inefficient and non-automated.
% totally depend on professional knowledge.
% In recent years, Ethereum phishing scams detection methods have focused on the detection of phishing addresses, 
% Features are manually extracted structure and statistical information of transactions network for integrated learning training. Since statistical features usually need to be manually proposed by professionals familiar with the field and accumulated experience, such methods are inefficient and non-automated. 
The other one is to apply network representation learning to the Ethereum transaction network for mining deep features. Random walk\cite{wu2020phishers,yuan2020detecting} and graph neural network\cite{chen2020phishinggcn} are adopted to automatically learn representations from the Ethereum transaction network, which has made a very important breakthrough.
% xxx uses classical network representation learning algorithms such as node2vec to perform node learning representation, 
% However, the existing GRL-based detection methods have the following problems that hinder the performance of Ethereum phishing scam identification: 
% These two types of methods have gained important achievements, but there are still some remaining problems to be solved:
However, there are still two remaining problems: 
\textbf{(1) Lack of temporal transaction information.} The existing methods only adopt the last transaction record or even completely ignore these records, instead of taking temporal information of transaction records into consideration, which leads to the incomplete edge representations in the Ethereum transaction network.
% Both two types of methods rarely use temporal transaction information. They only use the last transaction between a pair of addresses to represent their transaction relationship,or completely ignore the transaction records between the addresses and connect them with uninformed edges. These approaches waste a large number of existing temporal transactions records which leads to incomplete modeling.
\textbf{(2) Weak node representation.} Only statistical and structural features extracted from transaction records are considered as the node representation, while trading features referring to the contextual information of transaction records are ignored totally. In summary, the lack of temporal transaction information and weak node representation finally cause the unsatisfactory performance of Ethereum phishing addresses detection.
To address the above challenges, in this paper, we propose \textbf{T}emporal \textbf{T}ransaction \textbf{A}ggregation \textbf{G}raph \textbf{N}etwork (TTAGN) to enhance phishing scams detection on Ethereum by effectively utilizing transaction temporal information. We first build a large-scale Ethereum multilateral directed transaction network graph, in which a node is a unique address and a directed edge refers to a transaction between two addresses, and obtain the nodes' basic statistical features. We design three modules to generate node representations by graph mining. In detail, the graph is fed to the \emph{Temporal Edge Representation} module which fully models and mines temporal information of Ethereum transaction records between transaction nodes to generate the edge representations. In the \emph{Edge2node} module, the edge representations around the node are aggregated to fuse topological interactive relationships into its representation, also named as trading features, which enriches the characteristics of the nodes. We also extract the common structural features in the \emph{Structural Enhancement} module and further combine statistical, structural and trading features to generate the final node representation. Finally, the obtained node representations are fed into the classifier to identify phishing nodes.
% Evaluated on the real-world Ethereum phishing scams dataset, our TTAGN outperforms the state-of-the-art methods.
% We perform extensive experiments to corroborate the effectiveness of our approach. The experimental results demonstrate the superior performance of TTAGN. 
% extensive experiments results demonstrate the superior performance of TTAGN and corroborate the effectiveness of our approach. 
Extensive experiments are conducted on real-world datasets to verify the effectiveness of TTAGN.
% the proposed approach.
% First, combining the transaction data and the labeled phishing addresses obtained from Etherscan, we build a large-scale Ethereum multilateral directed transaction network where the nodes are classified into labeled phishing and unlabeled addresses, and the edges present the transaction between the addresses. Second, to extract features from the large-scale Ethereum transaction network more accurately and efficiently, we propose a temporal transaction edge embedding module
% , using the sequence model LSTM to express multiple transactions between nodes in time series 
% to generate the edge embedding of the temporal interaction graph, thereby capturing the temporal pattern of interaction between nodes and make full use of transaction information. Then, we propose an edge2node module that applies an attention mechanism to give more weight to related edges, which can effectively alleviate the imbalance of data. The next, we design structural enhancement module, using auto encoder to extracting the node structure features of the transaction graph. Finally, we concatenate learned node embedding and adopt the LightGBM to classify the phishing and nonphishing addresses. 
% formulate the Ethereum phishing scam detection problem as a graph node classification task and
% To summarize, our main contributions is four-fold:

\textbf{Contributions.} Our contributions can be summarized as:
\begin{itemize}
    \item We propose a Temporal Transaction Aggregation Graph Network (TTAGN) to enhance the Ethereum phishing scams detection performance by combining trading, structural and statistical features.
    % \item All the transaction records between nodes in the Ethereum transaction graph are modeled to mine the temporal information and enrich the edge representation.
    \item All the directed transaction edges (records) between nodes (addresses) in the Ethereum transaction graph (network) are modeled to mine the temporal information and enrich the edge representation.
    \item The edge representations around each node are aggregated to fuse topological interactive relationships to generate the trading features.
    \item We conduct extensive experiments on real-world Ethereum phishing scam dataset and results show that TTAGN outperforms state-of-the-art methods on multiple metrics.
\end{itemize}

The remainder of the this paper is organized as follows. Section \ref{sec:relwork} summarizes the prior researches related to our work. Section \ref{sec:problem_difinition} introduces the problem statement of this paper. Section \ref{sec:methodology} highlights the overall
design of TTAGN and Section \ref{sec:experiments} illustrates the experiments. Section \ref{sec:conclusion} concludes the paper.

% Introduction to your project. Start from some common knowledge that most of the
% reader (in computer security) would have and then narrow down to the details of
% your project. Speak about why the project is important, and why the reader
% should care about it. Finally talk briefly about what are you have done (for
% final project), what you are planning to do (for proposal). Reader should get a
% good chunk of understanding about your project from this introduction section
% (\secref{sec:intro}).
% It's good to finish introduction section with a quick list of contributions. 

% \subsection{For project proposal.}\label{sec:proposal}
% The sections mentioned here is just for reference. You are free to change them
% as you find suitable. In particular for proposal, some of the sections such
% as~\secref{sec:eval} might not make much sense. You can skip that. 

%%% Local Variables:
%%% mode: latex
%%% TeX-master: "main"
%%% End:

%  LocalWords:  biometrics cryptographic parallelized lossy

%% file: related_work.tex
\section{Related Work}
\label{sec:relwork}
% Phishing scams detection on Ethereum is a new fraud scenario. In this section, we will introduce Ethereum phishing scams detection related methods. Next, we will present the related applications exploring network representation.
Phishing scams detection on Ethereum is a new fraud scenario. In this section, we first briefly review prior work on Ethereum phishing scams detection. Next, we review the network representation learning which is the core task of Ethereum phishing scams detection.
\subsection{Ethereum Phishing Scams Detection}
\label{sec:overview}
For the phishing scams detection problem on Ethereum, there are two main categories of existing methods. 

The former mainly employ shallow models such as traditional machine learning methods with dedicated feature engineering, focusing on statistical features. Chen et al.\cite{chen2020phishing} extracted 219-dimensional statistical features from the node's 1-order and 2-order neighbors, including the node's in-degree, out-degree, maximum transaction value, and so on. Then they used a LightGBM-based ensemble machine learning algorithm to identify phishing nodes.
% During the classification process, they sampled the data and extracted 219-dimensional statistical features. 

The latter applies some network embedding methods such as DeepWalk\cite{deepwalk}, Node2Vec\cite{node2vec}, and graph convolutional networks (GCN)\cite{GCN} to mine deep features. 
% Yuan et al.\cite{yuan2020detecting} directly use the Node2vec algorithm on the Ethereum transaction network to learn the representation of nodes. 
Wu et al.\cite{wu2020phishers} proposed Trans2Vec on the basis of Node2Vec\cite{node2vec}. The difference between Trans2Vec and Node2Vec is that the sampling process of Trans2Vec is not random, but biased based on the last transaction of the two nodes, which is more suitable for phishing detection on Ethereum. Chen et al.\cite{chen2020phishinggcn} designed E-GCN to detect phishing nodes, which is the first time GCN\cite{GCN} has been introduced in Ethereum phishing node detection. They extracted 8-dimensional statistical features and then used GCN to learn the structural characteristics of the transaction network.
%  GCN is achieved by a localized first-order approximation of spectral graph convolutions.

However, these works rarely use the temporal information of transaction behaviors, so they can not capture complete edge representations. Moreover, only manual-designed features are taken for the node representation, which further led to weak node representation capabilities of these detection methods.

\subsection{Network Representaion Learning}
\label{sec:overview}
% Our model aims to efficiently learn the representation of nodes from the large-scale transaction information and network structure, the existing network representation learning methods can give us some inspiration. 
According to a survey\cite{survey2018graph}, network representation learning (i.e., graph embedding or network embedding) methods can be summarized into three categories: based on (1) Factorization, (2) Random Walk, and (3) Deep Learning.

Factorization-based algorithm uses the connection information between nodes to construct various matrices (e.g., Laplacian matrix, adjacency matrix, and Katz similarity matrix), and then factorize the above matrix to obtain embeddings. The models associated with factorization are, for example, Locally Linear Embedding (LLE)\cite{LLE}, Laplacian Eigenmaps\cite{laplacian}, Graph Factorization\cite{ahmed2013distributed}, Learning Graph Representations with Global Structural Information (GraRep)\cite{cao2015grarep}, High-Order Proximity Preserved Embedding (HOPE)\cite{HOPE}.

Random walk-based algorithm utilizes walk to perceive the centrality and similarity of nodes. DeepWalk\cite{deepwalk} tries to maximize the co-occurrence probability of nodes in the window after obtaining the node sequence of random walk. 
% Locally Linear Embedding ( Learning Graph Representations with Global Structural Information ( High-Order Proximity Preserved Embedding ( Large-scale Information Network Embedding ( Structural Deep Network Embedding ( Graph Attention Networks (
% In these algorithms, taking DeepWalk
% \cite{deepwalk} as a typical example, it tries to maximize the co-occurrence probability of nodes in the window after obtaining the node sequence of random walk. 
% Here, "window" refers to a custom, continuous node series in a walk sequence.
% The most significant difference between DeepWalk and Node2Vec\cite{node2vec} is that the walking strategy. 
As for Node2Vec\cite{node2vec}, in the first stage of generating nodes’ corpus, the walking decision is more flexible than DeepWalk, but the time consumption increases greatly. Different from DeepWalk, Large-scale Information Network Embedding (LINE)\cite{line} aims to generate neighbors rather than nodes on a path based on current nodes. 
% LINE defines two probability distributions for each pair of nodes, corresponding to the first- and second-order proximities and aims to minimize their Kullback-Leibler (KL) divergence\cite{kl}. 
% Although LINE is not based on a random walk method, it is an improvement based on DeepWalk, so we classify it into the same category as DeepWalk.
% So in the comparison of related work and experiments, we classify it into the same category as DeepWalk.

Deep learning-based method mainly uses deep neural networks to learn non-linear information in graphs. Structural Deep Network Embedding (SDNE)\cite{wang2016sdne} apply deep autoencoders to keep network proximities within 2-order. It uses a semi-supervised autoencoder to reconstruct the neighbor relationships of the nodes and uses a supervised approach to trim the results. 
% Deep Neural Networks for Learning Graph Representations (DNGR)\cite{cao2016dngr} uses a random surfing model to get a probabilistic co-occurrence matrix, and then the matrix is converted to get the result as the input of a stacked denoising autoencoder. 
% Related to deep neural networks, as well as GCN and Variational Graph Auto-Encoders (VGAE)\cite{kipf2016vgae} proposed by Kipf et al. VGAE uses GCN for encoding, then takes the inner product of the GCN matrix plus the  Kullback-Leibler (KL) divergence as the decoder. 
GraphSAGE\cite{hamilton2017graphsage} is an inductive GNN model based on a fixed sample number of the neighbor nodes and Graph Attention Networks (GAT)\cite{velivckovic2017gat} employs attention mechanism for neighbor aggregation.

We have selected representative works from three categories for comparison in the subsequent experimental part, which further highlights the effectiveness of our model.

%% file: problem_definition.tex
\section{Problem Definition}
\label{sec:problem_difinition}

In this paper, the Ethereum phishing scams detection task is phrased as a graph node classification problem.
% The principal aim of our work is to detect phishing scams on the Ethereum transaction network. We formulate the Ethereum phishing scams detection problem as a graph node classification task. 
% , and build a large-scale Ethereum multilateral directed transaction network in which data is obtained from authority agencies.
% We treat the transaction address as a node $v_{i}$, the transaction as an edge $\mathcal{E}_{i}$, and the transaction direction, amount and time information as the attributes of the edge, and finally a graph $\mathcal{G}$ is formed. Let the Ethereum transaction network $\mathcal{G}=(\mathcal{V}, \mathcal{E})$, $\mathcal{V}=\left\{v_{1}, \ldots, v_{N}\right\}$ is a set of nodes, $\mathcal{E}=\left\{\mathcal{E}_{1}, \ldots, \mathcal{E}_{R}\right\}$ is the edge set of $R$ relations. $\mathcal{G}_{L}=(\mathcal{V}, \mathcal{E}, X, C)$ is a partially labeled network, with edge attributes $X \in \mathbb{R}^{|\mathcal{E}| \times S}$ where $S$ is the size of the feature space for each edge, and $C \in \mathbb{R}^{|\mathcal{V}| \times |\mathcal{Y}|}$ where $\mathcal{Y}$ is the set of labels. 
Let the partially labeled Ethereum transaction network $\mathcal{G}_{L}=(\mathcal{V}, \mathcal{E}, X, C)$, we treat the transaction address as a node $v_{i}$, $\mathcal{V}=\left\{v_{1}, \ldots, v_{N}\right\}$ is a set of addresses. The transaction as an edge $\mathcal{E}_{i}$, $\mathcal{E}=\left\{\mathcal{E}_{1}, \ldots, \mathcal{E}_{R}\right\}$ is the transaction set. The transaction direction, amount and time information as the edge attributes $X \in \mathbb{R}^{|\mathcal{E}| \times S}$ where $S$ is the size of the feature space for each edge, and $C \in \mathbb{R}^{|\mathcal{V}| \times |\mathcal{Y}|}$ where $\mathcal{Y}$ is the set of labels. 
% Let the partially labeled Ethereum transaction network $\mathcal{G}_{L}=(\mathcal{V}, \mathcal{E}, X, C)$, we treat the transaction address as a node $v_{i}$, $\mathcal{V}=\left\{v_{1}, \ldots, v_{N}\right\}$ is a set of addresses. The transaction as an edge $\mathcal{E}_{i}$, $\mathcal{E}=\left\{\mathcal{E}_{1}, \ldots, \mathcal{E}_{R}\right\}$ is the transaction set of $R$ relations. The transaction direction, amount and time information as the edge attributes $X \in \mathbb{R}^{|\mathcal{E}| \times S}$ where $S$ is the size of the feature space for each edge, and $C \in \mathbb{R}^{|\mathcal{V}| \times |\mathcal{Y}|}$ where $\mathcal{Y}$ is the set of labels. 
% Some nodes in the graph are labeled, and we hope to learn the representation of all nodes through labeled nodes to further identify the unlabeled nodes.
% In the Ethereum transaction network, each edge contains two critical attributes, namely, transaction amount and timestamp. For the scenario of phishing address identification, $\mathcal{Y}$ contains two labels, i.e., $+1$ for phishing node and $-1$ for normal samples. 
The goal of our model is to efficiently learn the representation of nodes from the known large-scale transaction network information, so we learn the embeddings of all nodes $X_{E} \in \mathbb{R}^{|V| \times d}$, where $d$ is the number of dimensions for feature representation.

% The principal aim of this work is to detect phishing scams on an extremely large-scale Ethereum network. Because of the large-scale of the network and the imbalance of data labels, we propose a biased network embedding algorithm, which incorporates the transaction amount and timestamp of each edge to better capture the information from the Ethereum transaction network. The goal of the network embedding algorithm is to learn the embeddings of all nodes $X_{E} \in \mathbb{R}^{|V| \times d}$, where $d$ is the number of dimensions for feature representation. These obtained node embeddings can be used as feature inputs for the downstream classification task. Fig. 3 gives a simple illustration of the embedding procedure on the Ethereum transaction network.

% efficiently learn the representation of nodes from the known large-scale transaction information

%% file: methodology.tex
%%%%%%%%%%%%%%%%%%%%%%%%%%%%%%%%%%%%%%%%%%%%%%%%%%%%%%%%%%%%%%%%%%%%%%%%%%%%%%%%
\section{Design of TTAGN}
\label{sec:methodology}
\begin{figure*}[htbp]
\centering
\includegraphics[width = 1\linewidth]{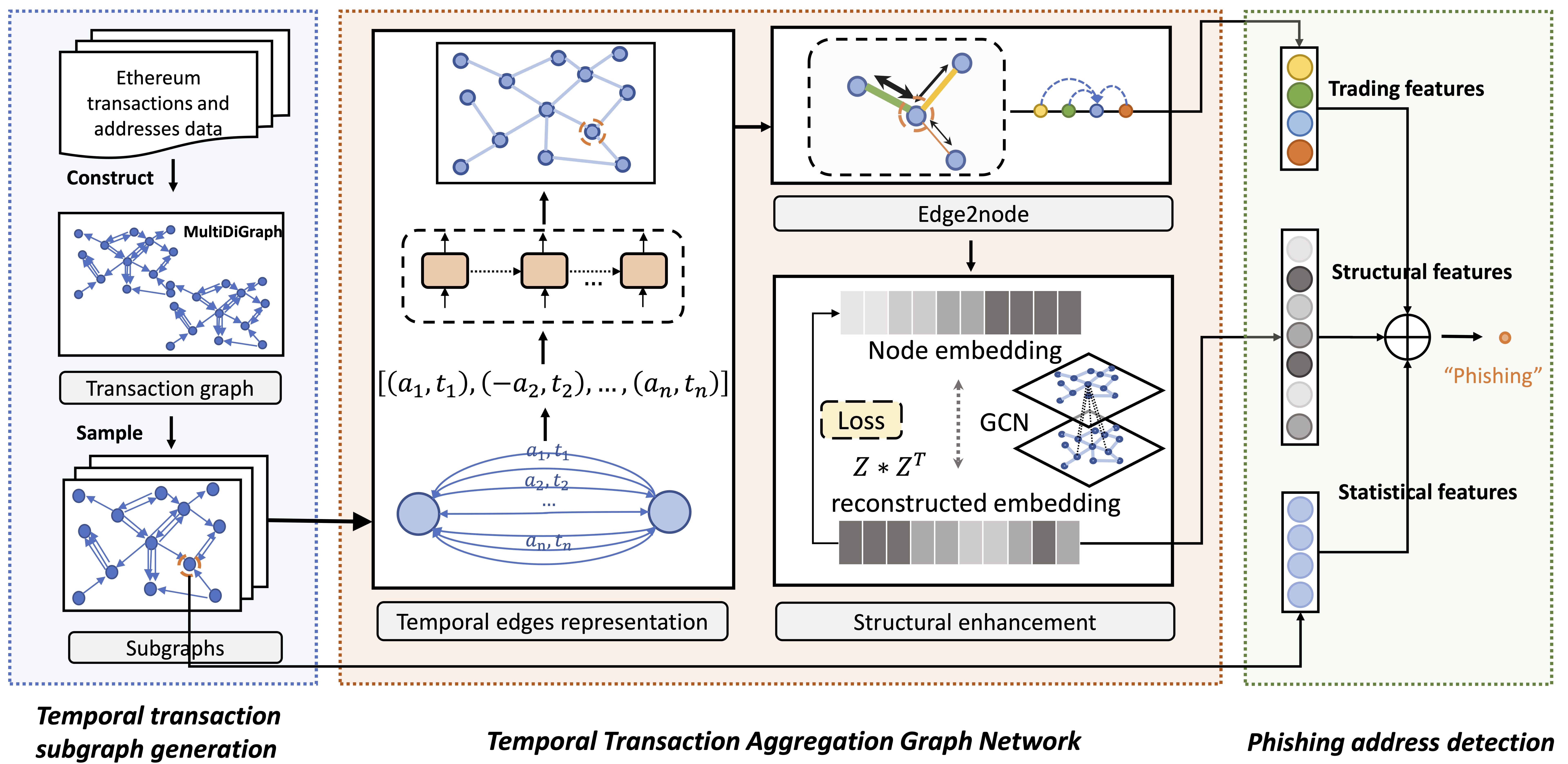}
\caption{The overall architecture of TTAGN. TTAGN inputs different sizes of Ethereum transaction subgraphs to learn their nodes' representation. The temporal edges representation module model the temporal relationship of historical transaction records to construct the representation of edges, the edge2node module aggregates edge representations around the node to fuse topological interactive relationships into trading features, and the graph autoencoder further enhances the perception of node structure information. Finally, the outputs of several modules are combined as the final representation of the node, which feeds into the classifier to get the result.}
\label{framework}
\end{figure*}
% This section introduces our phishing scam detection system, called TTAGN, which is a three-step detection, including building transaction graphs, learning network embedding and phishing addresses detection. 
% % The architecture of TTAGN is shown in Figure \ref{framework}.

TTAGN enhances the representation of edges by modeling transaction temporal information, finally improving the identification of Ethereum phishing nodes. Specifically, TTAGN includes three modules, named building transaction graphs, learning network embedding and phishing addresses detection. 

\subsection{Temporal Transaction Graphs}
\label{subsec:transaction}
Based on a large amount of Ethereum transaction data obtained, we first build a large-scale Ethereum transaction Multiple edges Directed Graph (MultiDiGraph). 
% First, we crawl the Ethereum transaction addresses data labeled "phishing" from the Ethereum label cloud\footnote{https://etherscan.io/accounts/label/phish-hack}. Then, we take advantage of the provided API by Etherscan to obtain transactions between addresses. Because the transaction data of Ethereum is too large, we centered on the addresses marked as phishing accounts and used two layers’ Breadth First Search (BFS) to extract addresses data and transactions between them. Finally, we generated a large-scale Ethereum transaction Multiple edges Directed Graph (MultiDiGraph). 
In the transaction graph, we use nodes to represent Ethereum transaction addresses, and edges to represent transactions between addresses. Noting that the Ethereum transaction MultiDiGraph allows multiple directed edges between any pair of nodes, and each edge carries information of the transaction, such as the transaction amount in ETH (the unit of Ether) and the execution timestamp.
% The original Ethereum transaction MultiDiGraph we constructed contained 6,844,161 nodes and 208,847,593 edges. 
The scale of the original graph is huge, so we take the sampling step with a random walk. 
We randomly select a node from the graph to start walking, then randomly select its neighbor as the next node, and repeat this process until the number of nodes reaches our requirement. 
After that we obtain subgraphs of the scale we want.
% We collected subgraphs of different sizes and each size of subgraph was collected 5 times.

% The collected subgraph will perform meta-feature engineering, which refers to the extraction of basic features of the nodes. As mentioned earlier, the nodes here have no portrait information, and the feature engineering can only be implemented by the basic information in relationship. Because the core of this article is that edge embedding and attention convolutional network can better aggregate nodal and structural features of the Ethereum transaction network and optimize downstream tasks, so we will not explore feature engineering deeply. In general, our model can be used as a strengthening component of feature engineering. Finally, we extract the following meta-features for the nodes, which can be combined to form more complex features.

The collected subgraph will first perform feature engineering to prepare for the subsequent learning of the structural features of the node. Because of the anonymity of the blockchain platform, the node itself does not carry any attribute characteristics. So we extract the following 10-dimensional features as the attribute characteristics of the node. They are the node's total degree, out-degree, in degree, the sum of transactions amount, transfer out transaction amount, transfer in transaction amount, the total number of neighbors, the inverse of transaction frequency, the percentage of neighbors whose transactions are all zeros, and the number of transactions with the most frequent neighbors.

\subsection{Model Architecture}
TTAGN is a network representation framework to detect phishing addresses. As shown in Figure \ref{framework}, the architecture could be divided into three objectives: temporal edges representation, edge2node, structural enhancement.

\subsubsection{Temporal Edges Representation.}
% Different from the general graph network, the transaction network is characterized by multiple directed transaction edges between a pair of nodes. Each transaction edge between a pair of nodes carries the amount and timestamp of this transaction. However, the existing works either do not use the transaction edge information at all, or only take the last transaction between nodes as the edge representation, which totally ignored the temporal relationship of historical transaction records and leads to incomplete edge representation learning. Based on this, we propose a temporal transaction edge representation method, apply the sequence model LSTM\cite{LSTM} to characterize the multiple temporal transactions between a pair of nodes, and generate the edge representation of the transaction graph. Hence, we can capture the temporal pattern of interaction between nodes and enrich the representation of edges. The flow of the temporal transaction edge embedding module is shown in the Figure \ref{process_temporal}.

% The transactions between each pair of nodes are treated as a time series, sorted in ascending order by timestamp, and then be fed into the LSTM model. For the node pair $(u, v)$, we denote $\tilde{e}_{uv}$ as the edge embedding generated by the sequence model LTSM, where:
In this module, edge representations are generated from transactions interaction relationships between nodes. 

This module improves the detection effect by introducing transaction temporal information. Transaction information includes transaction direction, amount, time, etc., which will reflect the difference between phishing addresses and normal at the transaction level. Therefore, by introducing the transaction information, the nodes' representations are enhanced.
% Different from the general graph network, the transaction network is characterized by multiple directed transaction edges between a pair of nodes. Each transaction edge between a pair of nodes carries the amount and timestamp of this transaction. 
% Therefore, we consider fusing the transaction amount and time information into the edge representation through modeling. For transaction networks, the amount, time, and direction of transactions between nodes are very important to the judgment of node relationships, so we need to mine deeper into these transaction information. 

% Transaction information has two characteristics, one is temporal sequence, the other is variable length. 
However, there are two difficulties with using transaction information directly: (1) Sequential. Transactions are temporal and inherently sequential, this information needs to be incorporated into edge representations. (2) Variable length. The number of transactions between nodes is different, edge representations should include all valid information without causing information redundancy.
% However, the existing works either do not use the transaction edge information at all, or only take the last transaction between nodes as the edge representation, which totally ignored the temporal relationship of historical transaction records and leads to incomplete edge representation learning. 

\begin{figure*}[htbp]
\centering
\includegraphics[width = 1\linewidth]{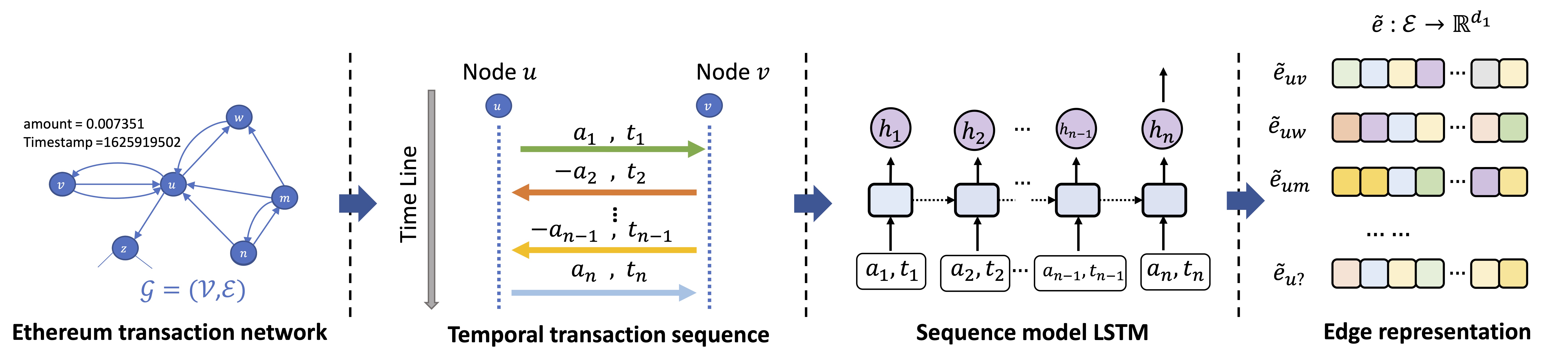}
\caption{Process of learning edge representation from the Ethereum transaction network.}
\label{process_temporal}
\end{figure*}

As for sequential, we apply the sequence model LSTM\cite{LSTM} to characterize the multiple temporal transactions and capture the temporal pattern of interaction between a pair of nodes. 
% and generate the edge representation of the transaction graph. 
% Hence, we can capture the temporal pattern of interaction between nodes and enrich the representation of edges. 
% The flow of the temporal transaction edge embedding module is shown in the Figure \ref{process_temporal}.
As Figure \ref{process_temporal} shows, the transactions between each pair of nodes are treated as a time series, sorted in ascending order by timestamp, and then be fed into the LSTM model. For the node pair $(u, v)$, we denote $\tilde{e}_{uv}$ as the edge embedding generated by the sequence model LTSM, where:
\begin{equation}
\begin{aligned}
\tilde{e}_{uv}&=LSTM\left (  \left [ e_{uv}^{1}, e_{uv}^{2}, \cdots,e_{uv}^{n} \right ]\right ) \\
&=LSTM\left (  \left [ \left (a_{uv}^{1} , t_{uv}^{1}\right ), \left (a_{uv}^{2} , t_{uv}^{2}\right ), \cdots,\left (a_{uv}^{n} , t_{uv}^{n}\right ) \right ]\right )
\end{aligned}
\end{equation}
% \begin{equation}
% e_{uv}^{i}=\left (a_{uv}^{i} , t_{uv}^{i}\right )
% \end{equation}
Among them, $a_{uv}^{i}$ represents the transaction amount of the $i$-th transaction with direction between nodes $u$ and $v$. The plus or minus of $a_{uv}^{i}$  represents the direction of this transaction, if the node transfers ETH to other nodes, $a$ is positive, else, the value is negative. $t_{uv}^{n}$ represents the transaction timestamp of the $n$-th transaction between nodes $u$ and $v$. 
As for variable length, we realize the variable-length input of LSTM, further make full use of the temporal transaction records.

Combining the above two points, 
% we convert multiple directed transaction edges to an undirected edge, and remodel the directed graph with multiple edges to an undirected graph with edge representation. We use directional transaction information to fully mine the transaction patterns between nodes, although eventually forming an undirected graph, edge embedding contains the directions information of all transactions. 
% It is worth noting that the transaction relationship (i.e., edges in the graph) is directed, but the neighbor relationship is undirected. Undirected graphs are beneficial to subsequent structural feature extraction and comparison experiments. 
% In the temporal edges representation module, 
we captured the temporal relationship of historical transaction records, generated effective edge representations, and simplified the complex graph structure (convert MultiDiGraph to undirected graph with edge representations), which is helpful for the subsequent node classification work.

% After getting the edge representation, we convert multiple directed transaction edges to an undirected edge, and remodel the directed graph with multiple edges to an undirected graph with edge representation. We use directional transaction information to fully mine the transaction patterns between nodes, although eventually forming an undirected graph, edge embedding contains the directions information of all transactions. It is worth noting that the transaction relationship (i.e., edges in the graph) is directed, but the neighbor relationship is undirected. Undirected graphs are beneficial to subsequent structural feature extraction and comparison experiments. In the temporal edges representation module, we captured the temporal relationship of historical transaction records, extracted effective edge representations, and simplified the complex graph structure, which is helpful for the subsequent node classification work.

\begin{figure}[htbp]
\centering
\includegraphics[width = 0.8\linewidth]{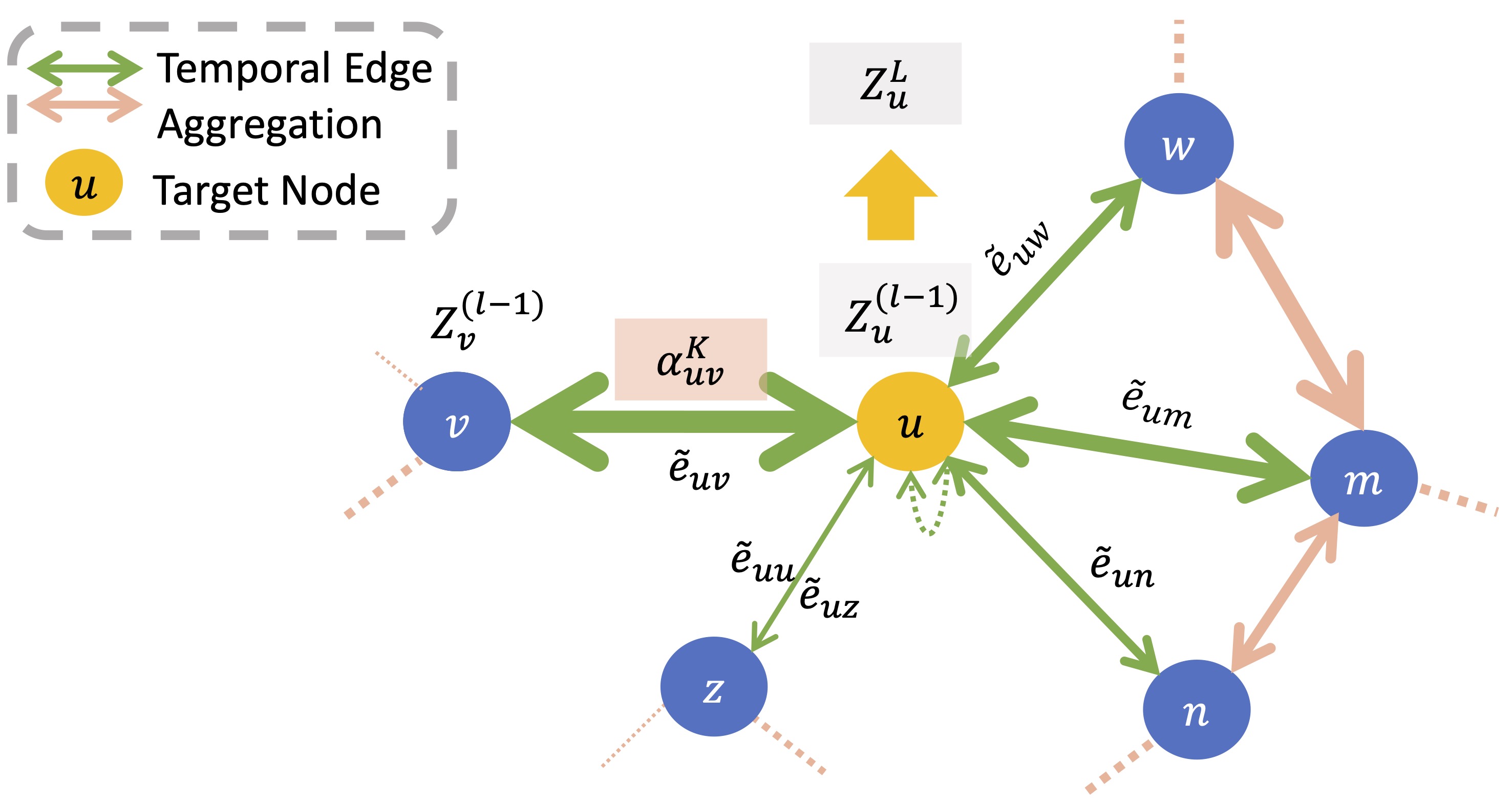}
\caption{Illustration of the edge2node module. 
% For a pair of nodes, fine-grained temporal edge embedding is obtained from the Temporal Edges Representation module. In edge2node module, we apply multi-head attention mechanism to catch similar transaction behaviors. The learned node embedding will be recursively aggregated with node attributes and fed to the structure enhancement module for further learning.
}
\label{attention}
\end{figure}

\subsubsection{Edge2node.} In this module, nodes representations are enriched by biased aggregation of edge representations with temporal transaction information to the nodes. 

In the transaction network on Ethereum, the node itself does not carry information, and only manual-designed features are not comprehensive which leads to weak node representation. 
Each Ethereum transaction node usually interacts with multiple nodes at the same time, and is also connected with multiple transaction edge representations. We need to fuse its interaction with all other nodes into its representation.
Meanwhile, different interaction have different effects on the node representation. 
% In fact, the transaction information between nodes can reflect their relationship, and different transactions have different effects on the node relationship. 
% For example, the larger the transaction amount, the later the transaction time will have a greater impact on the node relationship. 

To solve these problems, we aggregated the edge representations around each node to fuse topological interactive relationships. Moreover,
% The ideal situation is that the proportion of neighbor nodes related to the node type is large, and the proportion of neighbor nodes that are not related to it is small. 
we adopt Attention~\cite{vaswani2017attention} with multiple levels mechanism to catch similar transaction behaviors, and finally generate the trading features. Figure \ref{attention} shows the main steps of edge2node.
For each node of input transaction graphs, the edge2node learns the weights of adjacent edges and aggregates them to get the expressive node representation. 
Given $N_{u}$ denotes the adjacent edges of node $u$ and edge $v \in N_{u}$, the importance node-edge pair $\left \langle u,v \right \rangle$ can be formulated as follows:
% \begin{equation}
% e_{i j}=a\left(W \overrightarrow{h_{i}}, W \overrightarrow{h_{j}}\right)
% \end{equation}
% \begin{equation}
% \alpha_{i j}=softmax\left(e_{i j}\right)=\frac{\exp \left(e_{i j}\right)}{\sum_{k \in \mathcal{N}_{i}} \exp \left(e_{i k}\right)}
% \end{equation}
% \begin{equation}
% \alpha_{i j}=\frac{\exp \left(LeakyReLU \left(\vec{a}^{T}\left[W \overrightarrow{h_{i}} \| W \overrightarrow{h_{j}}\right]\right)\right)}{\sum_{k \in \mathcal{N}_{i}} \exp \left(LeakyReLU\left(\vec{a}^{T}\left[W \overrightarrow{h_{i}} \| W \overrightarrow{h_{j}}\right]\right)\right)}
% \end{equation}
\begin{equation}
\begin{gathered}
e_{u v}^{\Phi}=\sigma\left(a_{\Phi}^{T} \cdot\left[h_{u} \| h_{v}\right]\right) \\
\alpha_{u v}^{\Phi}=\operatorname{softmax}_{\mathrm{v}}\left(e_{u v}^{\Phi}\right)=\frac{\exp \left(e_{u v}^{\Phi}\right)}{\sum_{k \in N_{u}^{\Phi}} \exp \left(e_{u k}^{\Phi}\right)}
\end{gathered}
\end{equation}
where $h_{u}$ and $h_{v}$ are the features of node $u$ and edge $v$, $a_{\Phi}$ is the attention parametrize matrix for transaction graph $\Phi$, $\sigma$ denotes the activation function, and $\left |  \right |$  denotes the concatenate operation.
% A larger weight coefficient $\alpha_{u v}^{\Phi}$ indicates matching similar neighbor data between node $u$ and edge $v$, which contributes to the trading features learning and classification task.

Then, the node $u$ trading features can be obtained by aggregating all edge neighbor attributes with the corresponding coefficients as follows:
\begin{equation}
z_{u}^{\Phi}=\|_{k=1}^{K} \sigma\left(\sum_{v \in N_{u}^{\Phi}} \alpha_{u v}^{k} \cdot h_{v}\right)
\end{equation}
where $z_{u}^{\Phi}$ is the learned trading features of node $u$ for the transaction graph $\Phi$, $K$ is the head number using the multi-head attention mechanism\cite{vaswani2017attention}. 

% The learned trading features will be concatenated with statistical features and fed into the structure enhancement module for further structural learning.

\subsubsection{Structural Enhancement.}
% The above two modules focus more on extracting effective transaction features. In order to obtain a comprehensive node representation, in this module, we pay more attention to extracting the node structure features of the transaction graph. Analogous to the idea of auto encoder, we reconstruct the relationship between the nodes of the transaction graph. We combine the trading features obtained from edge2node with statistical features as node embedding, and input them into the GCN as the encoder to learn the structural features of the node.

% GCN is a semi-supervised graph embedding method that makes use of Laplace transform to make the node aggregate the features of higher-order neighbors. 
% % GCN carries out convolution operation in the spectral domain, and each operation can aggregate an additional layer of features. A single-layer GCN can process the information on first-order neighbors. Similar to CNN, GCN stacks multiple graph convolutional layers to extract high-level node representations. 
% The spectral convolution function is formulated as
In this module, structural features are obtained by reconstructing the transaction graph.

The above two modules focus more on extracting effective transaction features. 
% Modules \emph{Temporal Edges Representation} and \emph{Edge2node} model edge embedding and trading features respectively. However, the structural features of nodes are also very important. In order to fully explore structural features, 
% Unlike edge embedding and trading features, structural features represent the direct adjacency which are also meaningful. However, the transaction graph is partially labeled, which is not conducive to supervised learning.
In order to obtain a comprehensive node representation, in this module, we pay more attention to extracting the node structure features of the transaction graph. 
Analogous to the idea of Graph Auto-encoder\cite{kipf2016vgae}, 
% In order to fully explore structural features, 
we reconstruct the relationship between the nodes of the transaction graph. We combine the trading features obtained from edge2node with statistical features as node embedding, and input them into the GCN as the encoder to learn the structural features of the node.

% To address the above problem, we apply the Graph Auto-encoder\cite{kipf2016vgae} to learn structural features. 
The spectral convolution function is formulated as
\begin{equation}
H^{(l+1)}=\sigma\left(\tilde{D}^{-\frac{1}{2}} \tilde{A} \tilde{D}^{-\frac{1}{2}} H^{(l)} W^{(l)}\right)
\label{equation:GCN}
\end{equation}
where $I$ is the identity matrix, $\tilde{A} = A + I$ is the adjacency matrix $A$ with added self-connections $I$. $\tilde{D}$ is the degree matrix of $\tilde{A}$, and $ W^{(l)}$, $\sigma \left ( \cdot  \right )$ is the layer-specific trainable weight matrix and activation function, respectively. $H^{(l)} \in R^{n \times k}$ means the matrix of activation in $l$ layer, while $n$ and $k$ denote the number of nodes and output dimensions of layer $l$.  
% Take $H^{(0)}= X$ as input, every convolution operation would capture additional layer’s neighbor’s features. 
% If the objects of the first matrix multiplication are $A$ and $X$, then they are equivalent to nodes combine first order node neighbor features. 
% If $X$ is the node embedding processed by the edge2node in the previous stage, a layer of GCN It is equivalent to combining the features of the second-order edge neighbor, because the edge2node module has aggregated the first-order edge embedding of the node. 
% The more such multiplications, the more layers of information that are abstractly merged. However, on the one hand, because the neighbor matrix has no self-loop so that its own node information is neglected, on the other hand, the matrix is not normalized, and the difference in numerical dimensions will have a huge impact. We summarize these reasons and improve these problems one by one, such as the adjacency matrix added self-connections and normalized, to form the Equation (\ref{equation:GCN}).

The overall framework of the module can be defined as follows
\begin{equation}
\begin{gathered}
Z=GCN\left(X,A\right)
\\ 
\hat{A}=\sigma\left(ZZ^{T}\right)
\label{z}
\end{gathered}
\end{equation}
% Among them, $X$ is the input node embedding, $Z$ is the representation of all transaction nodes learned in the last layer of GCN, where GCN is used to encode and learn the attribute information and structure information of the graph at the same time. $ZZ^{T}$ is an operation that reconstructs the original graph structure with $Z$, the inner product of the vector is used to represent the adjacency relationship between nodes. $ZZ^{T}$ is essentially a decoding process, and $\hat{A}$ is the reconstructed adjacency matrix obtained after decoding. The reconstructed loss can be written as
Among them, $X$ is the input node embedding, $Z$ is the representation of all transaction nodes learned in the last layer of GCN.
% , where GCN is used to encode and learn the attribute information and structure information of the graph at the same time.
$ZZ^{T}$ is an operation that reconstructs the original graph structure with $Z$, 
% the inner product of the vector is used to represent the adjacency relationship between nodes. 
which is essentially a decoding process, and $\hat{A}$ is the reconstructed adjacency matrix obtained after decoding. The reconstructed loss can be written as
\begin{equation}
\mathcal{L}_{recon}=\frac{\|\hat{A}-A\|_{F}^{2}}{n}
\end{equation}
$\left \| \cdot  \right \|_{F}$ denotes the $l_{2}$-norm of a vector, By minimizing the reconstruction loss $\mathcal{L}_{recon}$, $Z$ will learn a more comprehensive node representation that includes the structural features of the transaction graph nodes.
% For a certain node, if its reconstruction decoder can approximate its original structural information, there is a relatively low probability that it’s an abnormal node.

\subsection{Phishing Addresses Detection}
% After utilizing the proposed TTAGN to obtain node embeddings, we use them as feature inputs for the task of phishing scam detection on Ethereum.
% After the above operations, we have obtained three types of features: trading features learned from temporal edge representation and edge2node modules, structural features learned from structural enhancement module, and statistical features obtained from nodes. We spliced them together as the complete representation of the node and input them into the selected LightGBM classifier for Ethereum phishing addresses classification. More details of LightGBM and three types of features can be found in the Appendix.
The task of this section is to classify nodes to distinguish between phishing nodes and normal nodes. After the above operations, we have obtained three types of features: trading features learned from \emph{Temporal Edge Representation} and \emph{Edge2node} modules, structural features learned from \emph{Structural Enhancement} module, and statistical features obtained from nodes. We splice them together as the complete representation of the node. On the basis of obtaining complete node representations, we need to learn the difference between fishing and normal node representations. So we input them into the classifier for Ethereum phishing addresses classification. 

There are many choices of classifiers, and in this article, we choose LightGBM\cite{lightgbm}, which is a new GBDT (Gradient Boosting Decision Tree) algorithm supporting efficient parallel training. The key concept behind GBDT is to iteratively train the weak classifier (decision tree) to get the optimal model. The model has the advantages of beneficial training effect and difficult over-fitting. 

% First, combining the collected transaction records and the tagged phishing addresses from authoritative websites, we built a large-scale Ethereum transaction network, which is a multilateral directed graph. Among them, the nodes are classified into phishing addresses and other addresses, the edges present transactions between each pair of addresses, and there are multiple directed edges between each pair of nodes. Secondly, to extract features from the Ethereum transaction network more accurately and effectively, we designed a novel network embedding framework TTAGN, which is divided into three parts: temporal edges representation, edge2node and structual enhancement. Finally, we adopt the tree model LightGBM to classify phishing and other addresses.

%% file: experiments.tex
%%%%%%%%%%%%%%%%%%%%%%%%%%%%%%%%%%%%%%%%%%%%%%%%%%%%%%%%%%%%%%%%%%%%%%%%%%%%%%%%
\section{Experiments}
\label{sec:experiments}
In this section, we perform empirical evaluations to demonstrate the effectiveness of the proposed TTAGN framework. Specifically, we aim to answer the following research questions:
\begin{itemize}
\item \textbf{RQ1}: How effective is the proposed approach TTAGN for detecting phishing addresses on the Etherum transaction network?
\item \textbf{RQ2}: How does each component of TTAGN (i.e., temporal edges representation, edge2node and structural enhancement) contribute to the final detection performance?
\item \textbf{RQ3}: How much will the performance of TTAGN change by providing different maximum temporal sequence lengths or different attention hidden sizes?
\end{itemize}

\subsection{Datasets}

% \textbf{Data Collection.}
\subsubsection{Data Collection.}
% One of the most important tasks in establishing a phishing scams identification model is to find enough phishing account examples. As discussed in Section \ref{subsec:transaction}, we crawled accounts labeled "phishing" from the label cloud of the authorized website Etherscan. As of July 2021, 4,932 addresses have been verified and considered to be phishing addresses. With these labeled nodes being the central nodes, we extract their first-order, second-order neighbors and the transactions between all of them through the API provided by Etherscan. Then we construct a large transaction network with multiple directed weighted edges. We choose the largest connected component to do the research, which includes 6,844,050 nodes and 208,847,461 edges, and the number of marked phishing nodes is 4931. We call these phishing addresses positive examples and the rest negative examples. In our experiment, we start from a random node and use random walks to get subgraphs with sizes of 30,000, 40,000, and 50,000 respectively, denoted as $D_{1}$, $D_{2}$, $D_{3}$. For each subgraph of different sizes, we sample five times to ensure the effectiveness of the performance. Detailed data information is shown in Table \ref{dataset}.  
We crawled accounts labeled "phishing" from the Ethereum label cloud of the authorized website Etherscan\footnote{https://etherscan.io}. As of July 2021, 4,932 addresses have been verified to be phishing addresses. With these labeled nodes being the central nodes, we extract their first-order, second-order neighbors and the transactions between all of them through the API provided by Etherscan. 
% Then we construct a large transaction network with multiple directed weighted edges. We choose the largest connected component to do the research, which includes 6,844,050 nodes and 208,847,461 edges, and the number of marked phishing nodes is 4931. 
Finally, we obtain 6,844,050 Ethereum addresses and 208,847,461 transaction records.
% We call these phishing addresses positive examples and the rest negative examples. 
The scale of the original graph is huge, so we sample with random walks to obtain subgraphs as our datasets with sizes of 30,000, 40,000, and 50,000 respectively, denoted as $D_{1}$, $D_{2}$, $D_{3}$. For each subgraph of different sizes, we sample five times to ensure the effectiveness of the performance. Detailed data information is shown in Table \ref{dataset}.

% \textbf{Data Cleaning.}
\subsubsection{Data Cleaning.}
After getting all the data, we found that the class is very imbalanced. 
% The class imbalance ratio, i.e., the ratio of the size of the majority class (negative examples) to minority class (positive examples), exceeds 26,000.
% % Given that some addresses are not phishing addresses, we recommend that some obvious negative examples (i.e., non-phishing addresses) be eliminated before model training in order to build a more effective model. Therefore, referring to \cite{chen2020phishing}, we 
% We refer to the data cleaning steps of \cite{chen2020phishing}, eliminate obvious non-phishing addresses to build a more effective model. 1) We ignore all transactions that appear before block height 2 million (i.e., 2016-08-02), because of all phishing addresses are active after this time, so the early Ethereum transactions before this timestamp were cleaned. 2) We eliminate addresses with less than 5 or more than 1,000 transaction records which may be wallets or other normal types of accounts\cite{chen2020phishing,wu2020phishers}, and we also did data analysis which proves that these addresses are not phishing nodes. After data cleaning, the average number of remaining nodes in each subgraph is 46930, 37194, and 27538 respectively.
We refer to the data cleaning steps of ~\cite{chen2020phishing}, eliminating obvious non-phishing addresses to build a more effective model. (1) We clean all transactions that appear before timestamp 2016-08-02 because all phishing addresses are active after this time; (2) We eliminate addresses with less than 5 or more than 1,000 transaction records which may be wallets or other normal types of accounts~\cite{chen2020phishing,wu2020phishers}, and we also did data analysis which proves that these addresses are not phishing nodes. After data cleaning, the average number of remaining nodes in each subgraph is 46930, 37194, and 27538 respectively. In the final classification task, we set 80\% of the total data as training data and the rest as test data.

Finally, each subgraph is embedded through TTAGN to obtain the nodes' representations for the downstream classification task. In the final classification task, we set 80\% of the total data as training data and the rest as test data\cite{chen2020phishinggcn}.
\begin{table}[]
\caption{Statistics of evaluation datasets. Labeled represents the number of labeled nodes in the dataset, and each number is the average calculated by five subgraphs.}
\centering
\setlength{\tabcolsep}{0.6mm}{
\begin{center}
% \resizebox{!}{30pt}{
\begin{tabular}{c|cclc}
\toprule[1pt]
% \hline
\textbf{Dataset}  &  \textbf{\#Total Nodes} &  \textbf{\#Labeled} & \multicolumn{1}{c}{\textbf{\#Edges}}  &  \textbf{\#Average Degree} \\ 
\midrule
$D_{1}$ & 30000        & 108             & 25048388 & 834.9741         \\ \midrule
$D_{2}$ & 40000        & 139             & 27481082 & 687.0442         \\ \midrule
$D_{3}$ & 50000        & 170             & 29854251 & 590.8691         \\ 
% \hline
\bottomrule[1pt]
\end{tabular}
% }
\end{center}
}
\label{dataset}
\vspace{-0.7cm}
\end{table}

\subsection{Experimental Setup}
% \textbf{Comparison Methods.}
\subsubsection{Comparison Methods.} 
We compare our proposed TTAGN framework with four categories of Ethereum phishing scams detection methods, including (1) Feature-based methods where only the node attributes are considered\cite{chen2020phishing}, (2) Factorization-based network embedding methods\cite{LLE}, and (3) Random walk-based network embedding methods (i.e., DeepWalk\cite{deepwalk}, Node2Vec\cite{node2vec}, and LINE\cite{line}) where both topological information and node attributes are involved. In addition, we also use some of the popular (4) Deep learning-based network representation methods (SDNE\cite{wang2016sdne}, E-GCN\cite{chen2020phishinggcn}, GraphSage\cite{hamilton2017graphsage} and GAT\cite{velivckovic2017gat}) to learn nodes representations to compare with the representations learned by our method. 
% Details of these four types of comparison baseline methods are described in Section \ref{sec:relwork}.

\begin{itemize}
    \item \textbf{Features only}~\cite{chen2020phishing} are 219-dimensional statistical features from the node's 1-order and 2-order neighbors.
    \item \textbf{LLE}~\cite{LLE} factorizes the constructed matrix which uses the connection information to obtain embeddings. 
    % It is a non-linear dimensionality reduction algorithm, which can make the data after dimensionality reduction better maintain the original manifold structure.
    % \item \textbf{GraRep}~\cite{cao2015grarep}: Following the idea of matrix decomposition, GraRep analyzes that the information depicted by different k-steps (the number of steps in random walk) is different. Compared with a series of NRL algorithms represented by LINE, GraRep can better capture the relationship between distant nodes.
    % \item \textbf{HOPE}~\cite{HOPE}: The High-Order Proximity Preserved Embedding algorithm is proposed to learn the asymmetric transitivity in directed graphs, and the measurement of the asymmetric transitivity of the algorithm is observable.
    % \item \textbf{TADW}~\cite{yang2015TADW}: In addition to the connection information in the graph network, there are quite a lot of nodes with additional attribute information (such as text information). TADW considers and embeds the text information of the nodes in the graph network into the final low-dimensional distributed expression at the same time.
    \item \textbf{DeepWalk}~\cite{deepwalk} tries to maximize the co-occurrence probability of nodes in the window after obtaining the node sequence of random walk. 
    % This is the pioneering work to learn node representations via simulating unbiased random walks. DeepWalk 
    % samples truncated random walks to learn latent node embedding based on the Skip-Gram model, it treats random walks as the equivalent of sentences. 
    % After the process of node sampling, it learns node embeddings by predicting each node’s neighborhood.
    \item \textbf{Node2Vec}~\cite{node2vec} defines a more flexible notion of a node’s neighborhood and exploits a biased random walk to encode both local and global network structures.
    \item \textbf{LINE}~\cite{line} learns a low-dimensional embedding via preserving the first-order and second-order closeness of nodes.
    % LINE defines the first-order similarity and second-order similarity, representing whether nodes are directly connected and whether they share common neighbors. 
    % Different from DeepWalk, LINE aims to generate neighbors rather than nodes on a path based on current nodes. LINE seeks to learn a low-dimensional embedding in order to preserve the first- and second-order closeness of nodes.
    \item \textbf{SDNE}~\cite{wang2016sdne} uses semi-supervised autoencoder to reconstruct the neighbor relationships and supervised approaches to trim the results.
    % makes use of deep autoencoders to keep network proximities within second order. It 
    % Graph convolution network achieved by a localized first-order approximation of spectral graph convolutions.
    \item \textbf{GraphSAGE}~\cite{hamilton2017graphsage} is an inductive GNN model based on a fixed sample number of the neighbor nodes.
    \item \textbf{GAT}~\cite{velivckovic2017gat} employs attention mechanism for neighbor aggregation.
    \item \textbf{E-GCN}~\cite{chen2020phishinggcn} is the first time that Graph Neural Network has been applied to Ethereum phishing node detection. 
\end{itemize}

\begin{table*}[]
\caption{Performance comparison results w.r.t. AUC, Recall, Precision and F1-score on three datasets.}
\begin{tabular}{cc|cccc|cccc|cccc}
\toprule[1pt]
                            & Dataset             & \multicolumn{4}{c|}{$D_{1}$}     & \multicolumn{4}{c|}{$D_{2}$} & \multicolumn{4}{c}{$D_{3}$}                                                                                                              \\ \cmidrule{2-14} 
\multirow{-2}{*}{\textbf{Method}}    & Metric              & AUC   & Recall & Pre   & F1    & AUC  & Recall  & Pre  & F1 & AUC                           & Recall                        & Pre                                    & F1                            \\ \midrule
                            \multicolumn{1}{l}{Feature-based}       & Only Features & 0.807          & 0.669          & 0.699          & 0.684          & 0.778          & 0.524          & 0.723          & 0.607          & 0.733          & 0.575          & 0.686          & 0.633          \\ \midrule
\multicolumn{1}{l}{Factorization} & LLE           & 0.773          & 0.784          & 0.488          & 0.602          & 0.732          & 0.575          & 0.339          & 0.427          & 0.753          & 0.455          & 0.422          & 0.438          \\ \midrule
\multirow{3}{*}{Random walk}      & Deep Walk     & 0.790          & 0.499          & 0.755          & 0.601          & 0.742          & 0.515          & 0.398          & 0.449          & 0.733          & 0.367          & 0.632          & 0.464          \\
                                  & Node2Vec      & 0.602          & 0.414          & 0.550          & 0.472          & 0.717          & 0.519          & 0.475          & 0.496          & 0.826          & 0.735          & 0.434          & 0.545          \\
                                  & LINE          & 0.813          & 0.736          & 0.624          & 0.675          & 0.797          & 0.650          & 0.676          & 0.662          & 0.802          & 0.655          & 0.611          & 0.632          \\ \midrule
\multirow{4}{*}{Deep learning}    & SDNE          & 0.720          & 0.838          & 0.360          & 0.504          & 0.729          & 0.613          & 0.320          & 0.421          & 0.739          & 0.717          & 0.334          & 0.456          \\
                                  & E-GCN  & 0.722          & 0.615          & 0.607          & 0.698          & 0.806          & 0.761          & 0.762          & 0.761          & 0.765          & 0.703          & 0.626          & 0.662          \\
                                  & GAT           & 0.764          & 0.622          & 0.498          & 0.553          & 0.812          & 0.665          & 0.580          & 0.620          & 0.828          & 0.682          & 0.556          & 0.643          \\
                                  & GraphSAGE     & 0.838          & 0.675          & \textbf{0.731} & 0.702          & 0.804          & 0.634          & 0.577          & 0.604          & 0.802          & 0.665          & 0.619          & 0.641          \\ \midrule
Ours                              & TTAGN         & \textbf{0.903} & \textbf{0.855} & 0.721          & \textbf{0.783} & \textbf{0.910} & \textbf{0.833} & \textbf{0.807} & \textbf{0.820} & \textbf{0.928} & \textbf{0.859} & \textbf{0.777} & \textbf{0.816}\\
\bottomrule[1pt]
\end{tabular}
\label{table:compared methods}
\end{table*}
\subsubsection{Evaluation Metrics.}
In this paper, we use the following four metrics to have a comprehensive evaluation of the performance of different methods in terms of Ethereum phishing scam detection: \textbf{(1) Area Under Curve (AUC).} The AUC metric is to calculate the area under the ROC curve formed by TPRs and FPRs with multiple thresholds, which is frequently used in binary classification tasks. \textbf{(2) Recall.} The recall rate means the percentage of known phishing nodes samples detected. \textbf{(3) Precision.} The precision rate means the percentage of real phishing nodes are in the accounts that are judged to be suspicious. \textbf{(4) F1-score.} F1-score is a comprehensive evaluation of the Precision and Recall score.
% AUC is the area under the ROC Curve.

% AUC score judges the overall effectiveness of the model, besides AUC, we value the Recall the most. Recall rate means the percentage of known negative samples detected. The number of suspicious nodes is small, and we prefer to recall these nodes. In daily life, we are more tolerant of risk warnings than being scammed. So, our goal is to identify as many phishing nodes as possible within a reasonable range.

%  \textbf{Implementation Details.}
 \subsubsection{Implementation Details.}
% The embedding size of all models is fixed to 10. For attention, we set the attention hidden size to 2 and the learning rate to 0.01. For GCN in our method, we set two layers with 0.001 learning rate. For the LightGBM model, an early stopping strategy is performed, the number of leaves and the learning rate of the LightGBM model are empirically fixed at 50 and 0.03, respectively. Due to the imbalance of the data, we upsample the minority class with a ratio of 50. For all the comparison methods, we select the hyper-parameters with the best performance on the validation set and report the results on the test data of the target network for a fair comparison\cite{collell2018simple}. Particularly, for all the network-based methods, the whole network structure and node attributes are accessible during training. Each comparison method's hyper-parameters can be found in Appendix.
The embedding size of all models is fixed to 10. For attention, we set the attention hidden size to 2 and the learning rate to 0.01. For GCN in our method, we set two layers with 0.001 learning rate. For DeepWalk and Node2Vec, the walk length, window size, the latter’s $p$ and $q$ are set to 20 and 4, 0.25, 0.4, respectively. For the LightGBM model, the number of leaves and the learning rate are empirically fixed at 50 and 0.03, respectively. Due to the imbalance of the data, we upsample the minority class with a ratio of 50. 
% For all the comparison methods, we select the hyper-parameters with the best performance on the validation set and report the results on the test data of the target network for a fair comparison.
For all the comparison methods, we set parameters based on their official implementations.

% \subsubsection{Notes.}
% Our model TTAGN identifies phishing nodes through the node's period of time transactions. Refer to \cite{chen2020phishinggcn}, in the process of model learning, the unlabeled node representation and the parameters of the classifier are learned by inputting the partially labeled graph, and finally recognizing the unlabeled node’s identity. In the actual application process, input an unlabeled real-time Ethereum transaction network into the trained model TTAGN to detect the nodes in it. Changes in the number of transactions and nodes do not affect the detection, because the trained LSTM module can accept variable-length transaction sequences, and the unsupervised modules can automatically learn graph topology and aggregate edge embedding from unlabeled data. Finally, input the obtained node representation into the trained classifier, we can identify the phishing nodes from the Ethereum transaction network.
%  and the metrics of F1-score with the threshold 60. Moreover
% we search for the number of layers in \{1,2,3,4,5,6\} and tune the learning rate in \{0.001,0.005,0.01,..., 0.05\}. 
% Please add the following required packages to your document preamble:
% \usepackage{multirow}
% Please add the following required packages to your document preamble:
% \usepackage{multirow}

\subsection{Effectiveness Results (RQ1)}
To answer RQ1, we evaluate the performance of all the compared methods in the task of phishing scams detection on Ethereum. The corresponding results are reported in Table \ref{table:compared methods}. We can draw the following conclusions:
% The compared methods can be further divided into three groups.

(1) In terms of the four evaluation metrics, our approach TTAGN outperforms all the other compared methods by a significant margin. Our method TTAGN achieves the best performance about 92.8\%AUC, 85.9\% Recall, 77.7\% precision and 81.6\% F1-score under $D_{3}$ dataset. The second best method is deep learning methods which reach the AUC exceeding 80\%. The performances of the random walk-based method and the factorization-based method are similar, and their indicators are both around 75\%. The worst performance is the feature-based method, and its Recall rate is very low, only about 55\%.
% outperforms state-of-the-art methods on multiple metrics and datasets.

(2) TTAGN has better node representation capability on large graphs. As the number of datasets nodes increases from 30,000 to 50,000, the gap between TTAGN and other comparison methods is further widened. Compared with the well-performing GraphSAGE method, the AUC difference between the two methods is 6.5\% on $D_{1}$ and 12.6\% on $D_{3}$. These results again demonstrate that TTAGN can better detect phishing nodes on large-scale transaction networks than other methods by fully mining temporal information of Ethereum transaction records between transaction nodes.
% learning the temporal transactions information.

(3) Compared with the feature-based methods, our four evaluation metrics are nearly 20\% higher than them. The performance of the feature-only method is the worst across all compared methods. When the dataset is small, its effect is better than the factorization-based method, but as the number of nodes increases, the information that the statistical features can learn is very limited. Apart from the lack of feature mining, it may be because of these methods' unawareness of the network structure and environment information that we obtain from the structural enhancement module. 

(4) As for the random walk-based methods, LINE performed the best, which is lower than our method 12.6\% AUC and 18.4\% F1-score on the $D_{3}$ dataset. LINE uses the deep excavation of proximity within the second-order, through it LINE can perceive the nearby information than Deep Walk and Node2Vec. However, this type of methods completely ignores the transaction records between the nodes, which lead to incomplete representation learning of nodes. In our method TTAGN, we model the temporal relationship of historical transaction records, which makes full use of transaction information and learns the effective edge representation.

(5) Network representation methods based on deep learning are our strong opponents, however, they are also not performing well. On the dataset $D_{3}$, our four evaluation metrics are nearly 10\% higher than it. As for GraphSAGE, it does not explore label distribution when sampling, thus they perform worse than GAT. GAT performs worse than our method TTAGN because in the biased aggregation step, GAT aggregates neighbors with statistical characteristics, while we use the edge2node module to aggregate the obtained edge representations to nodes. This approach enriches the characteristics of the nodes and strengthens the nodes' representation ability.

\subsection{Ablation Study (RQ2)}
To answer RQ2 
% we identify the three key steps of TTAGN, i.e., namely temporal edges representation, edge2node and structural enhancement. TTAGN adopts the Temporal Edge Representation module and the Edge2node module to model the temporal transaction information and enhance the node representation. We expect to demonstrate the effectiveness of our innovation by eliminating 
and validate the effectiveness of our innovation, we eliminate the Temporal Edge Representation module (i.e. TTAGN/t), the Edge2node module (i.e. TTAGN/e) and the Structural Enhancement module (i.e. TTAGN/s) respectively. 
% We implement TTAGN/t, TTAGN/e, and TTAGN/s for the transaction network representation learning ablation study. The three variants respectively remove the temporal edges representation module, edge2node module or structural enhancement module when learning the nodes' representations.
% In this part, we denote the TTAGN that removes the temporal edges representation module as TTAGN/t, removes the edge2node module as TTAGN/e and removes the structural enhancement module as TTAGN/s.

% \begin{figure}[]
% % \centering
%     \subfigure[P-R curves]{
%       \begin{minipage}[t]{0.5\textwidth}
%           \centering
%           \includegraphics[width = 0.5\linewidth]{bar_1.pdf}
%       \end{minipage}
%       }%
%     \subfigure[ROC curves]{
%       \begin{minipage}[t]{0.5\textwidth}
%       \centering
%       \includegraphics[width = 0.5\linewidth]{bar_2.pdf}
%       \end{minipage}
%       }%
% \caption{Precision-Recall curves and Receiver Operating Characteristic curves on datasets}
% \label{fig:ablation}
% \end{figure}

\begin{figure}[htbp]
    \centering  %居中
    \begin{subfigure}{4cm}
        \includegraphics[scale=0.18]{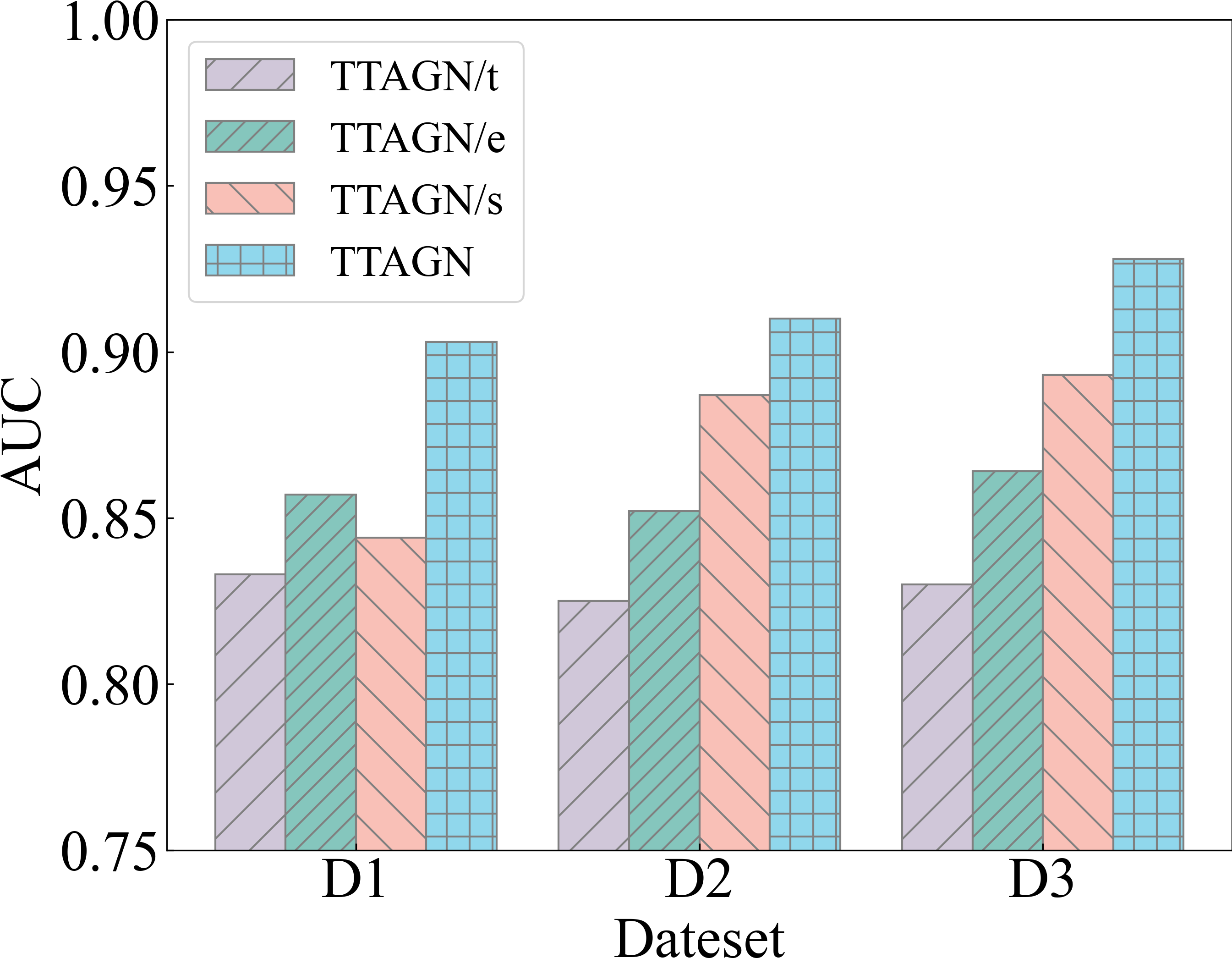}
        \caption{AUC Result}
    \end{subfigure}
    % \hskip2em
    % \hskip2em
    \begin{subfigure}{4cm}
        \includegraphics[scale=0.18]{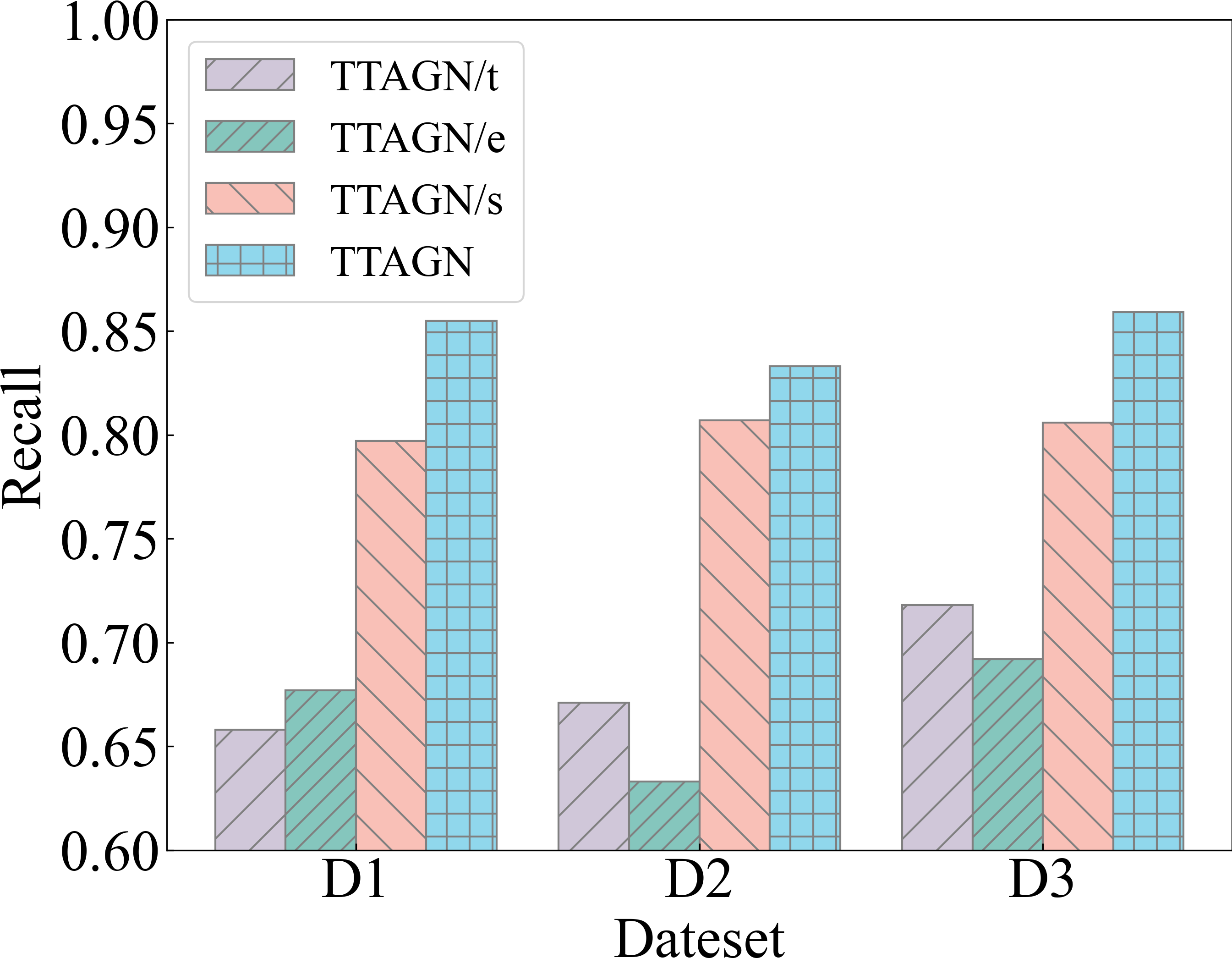}
        \caption{Recall Result}
    \end{subfigure}
    % \begin{subfigure}{.4\linewidth}
    %     \includegraphics[scale=0.7]{tt}
    %     \caption{tt}
    % \end{subfigure}
    % \caption{Several subfigures}
    \vspace{+0.5cm}
    \caption{AUC and Recall results of TTAGN and its variants.}    %大图名称
\label{fig:ablation}    %图片引用标记
\end{figure}

% \begin{figure}[htbp]
% \centering  %居中
% % \subfigure[AUC Result]{   %第一张子图
% \subfloat[AUC Result]{   %第一张子图
%     \begin{minipage}{4cm}
%     \centering    %子图居中
%     \includegraphics[scale=0.18]{pic/bar_1.png}  %以pic.jpg的0.5倍大小输出
%     \end{minipage}
% }
% % \subfigure[Recall Result]{ %第二张子图
% \subfloat[Recall Result]{ %第二张子图
%     \begin{minipage}{4cm}
%     \centering    %子图居中
%     \includegraphics[scale=0.18]{pic/bar_2.png}%以pic.jpg的0.5倍大小输出
%     \end{minipage}
% }
% \vspace{+0.5cm}
% \caption{AUC and Recall results of TTAGN and its variants.}    %大图名称
% \label{fig:ablation}    %图片引用标记
% \end{figure}

\begin{figure*}[htbp]
\centering  %居中
    \begin{subfigure}{0.235\textwidth}
        \includegraphics[scale=0.18]{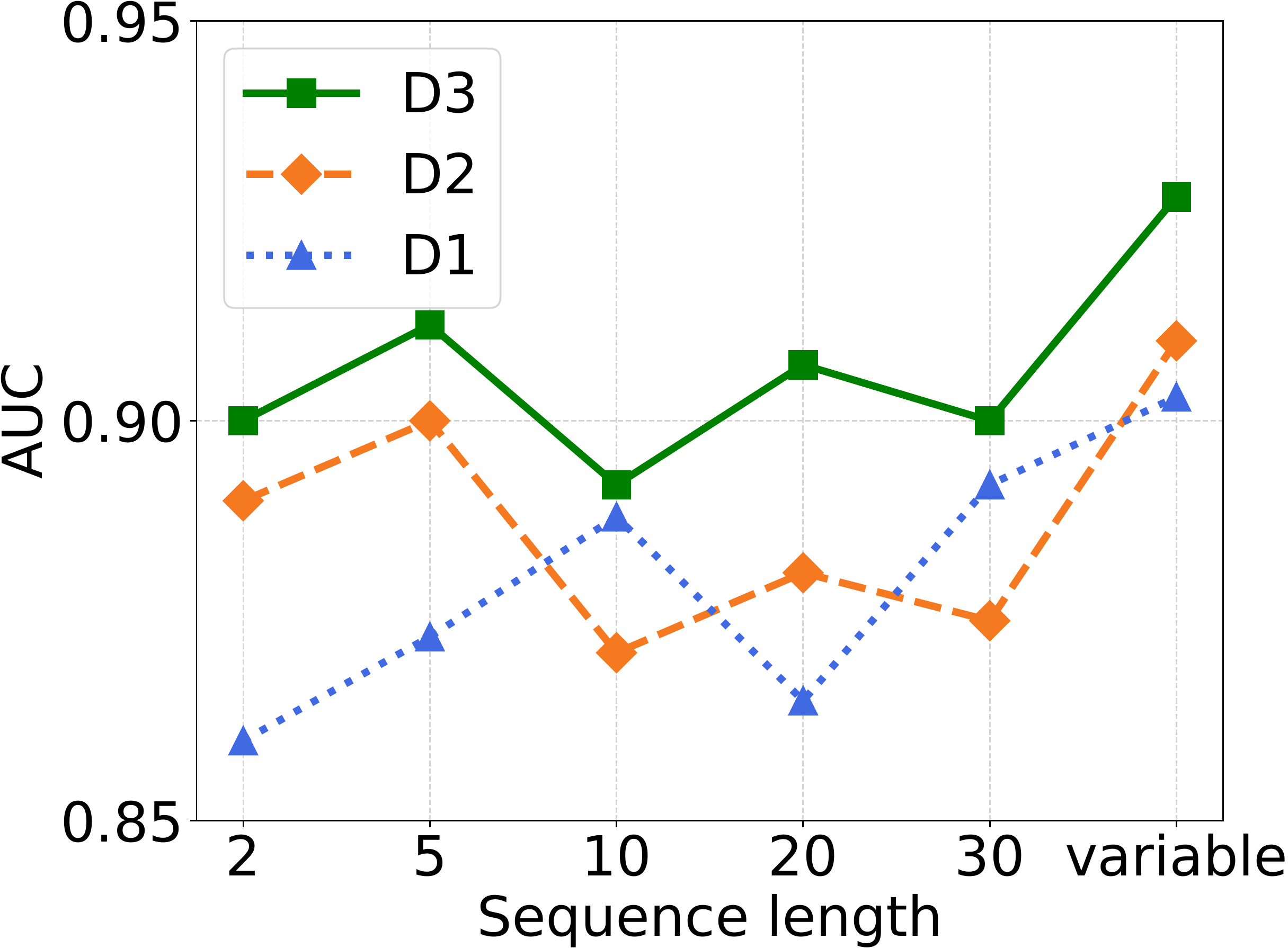}
        \caption{AUC}
    \end{subfigure}
    % \hskip2em
    % \hskip2em
    \begin{subfigure}{0.235\textwidth}
        \includegraphics[scale=0.18]{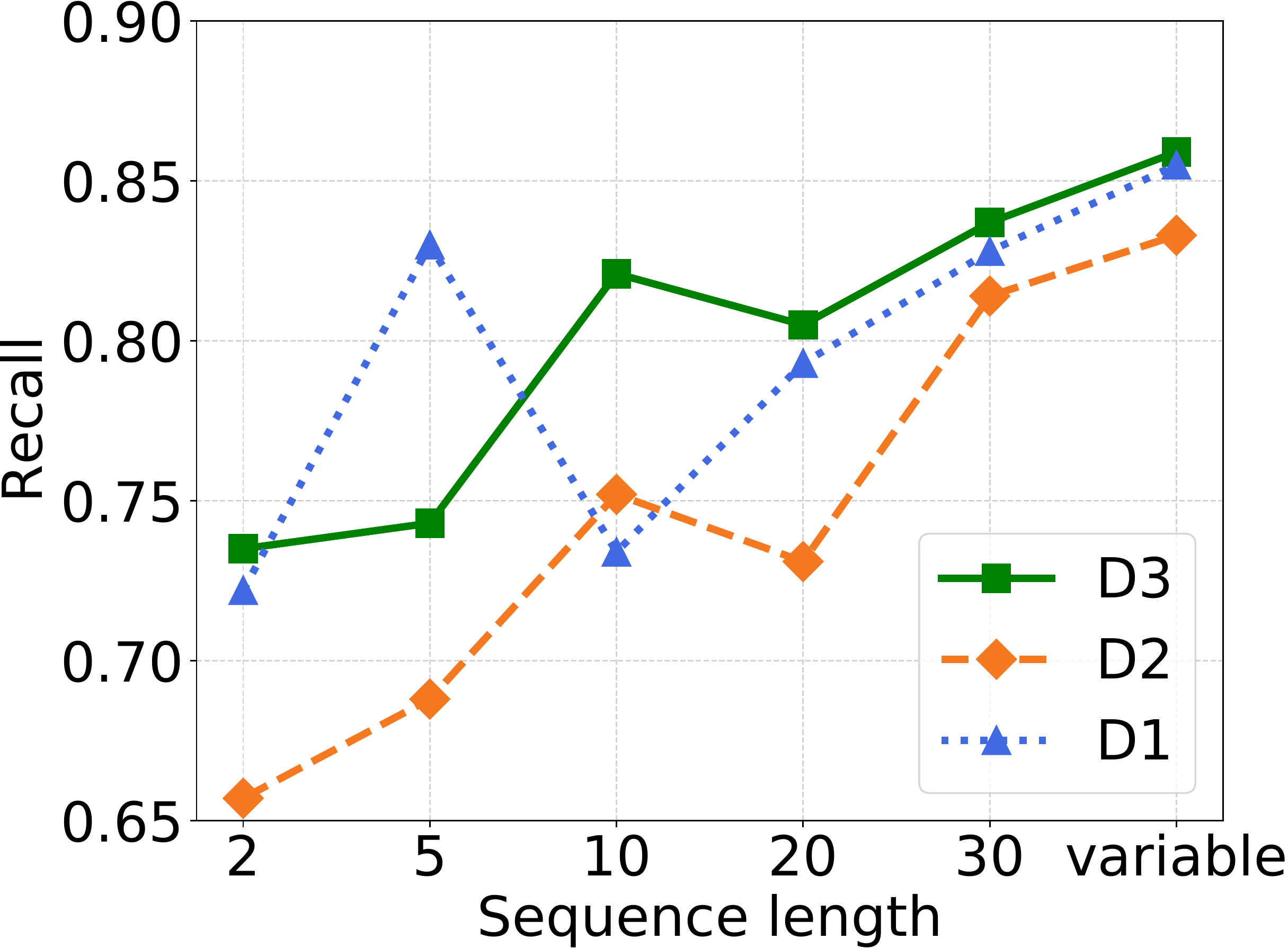}
        \caption{Recall}
    \end{subfigure}
    \begin{subfigure}{0.235\textwidth}
        \includegraphics[scale=0.18]{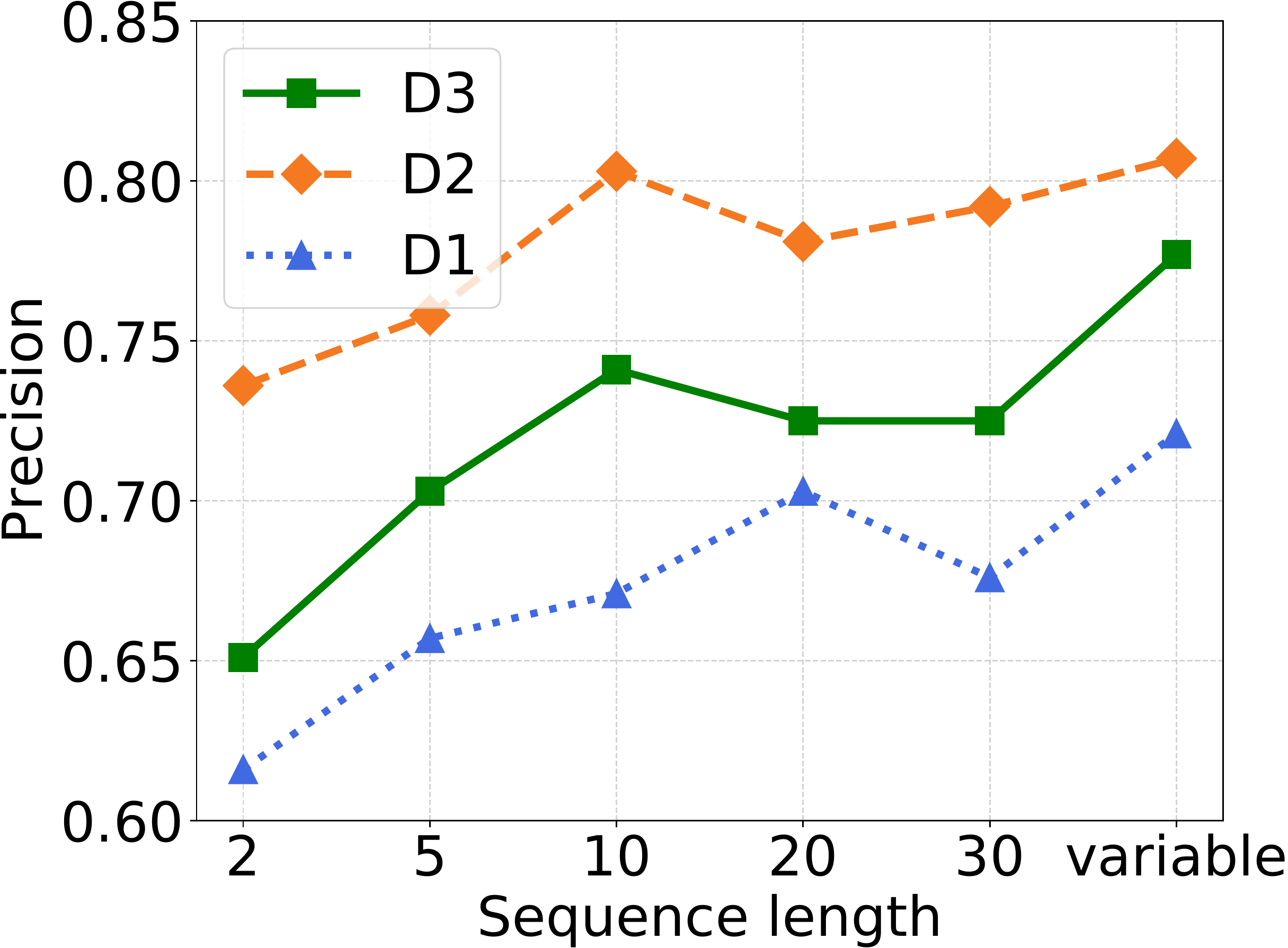}
        \caption{Precision}
    \end{subfigure}
    \begin{subfigure}{0.235\textwidth}
        \includegraphics[scale=0.18]{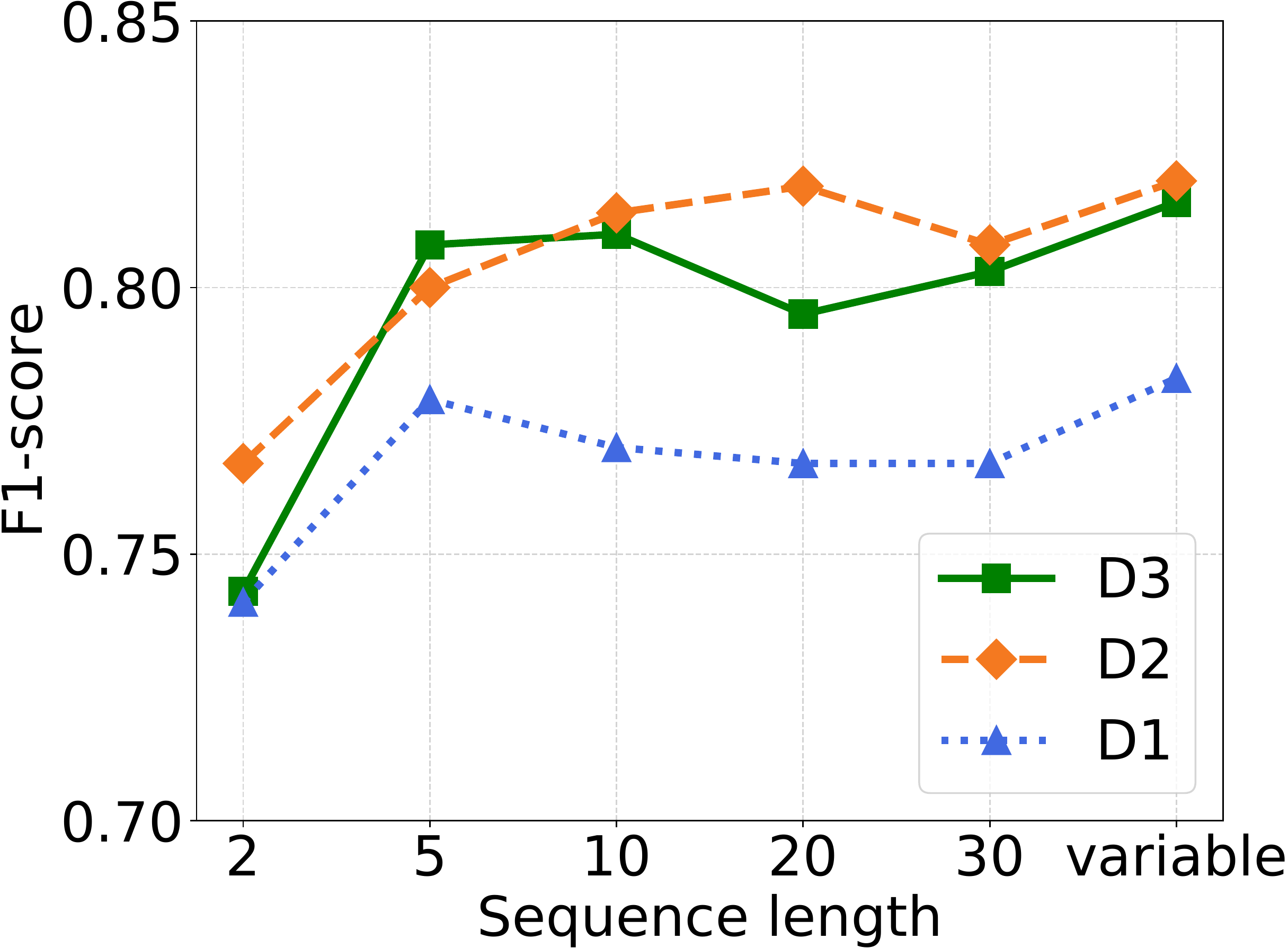}
        \caption{F1-score}
    \end{subfigure}
\vspace{+0.5cm}
\caption{Sensitivity analysis of TTAGN with different sequence lengths}    %大图名称
\label{fig:param_1}    %图片引用标记
\end{figure*}

% \begin{figure*}[htbp]
% \centering  %居中
% % \subfigure[AUC]{   %第一张子图
% \subfloat[AUC]{   %第一张子图
% \begin{minipage}[t]{0.235\textwidth}
% \centering    %子图居中
% \includegraphics[scale=0.18]{pic/param_1_auc.pdf}  %以pic.jpg的0.5倍大小输出
% \end{minipage}
% }
% % \subfigure[Recall]{
% \subfloat[Recall]{
% \begin{minipage}[t]{0.235\linewidth}
% \centering    %子图居中
% \includegraphics[scale=0.18]{pic/param_1_recall.pdf}%以pic.jpg的0.5倍大小输出
% \end{minipage}
% }
% % \subfigure[Precision]{ %第二张子图
% \subfloat[Precision]{ %第二张子图
% \begin{minipage}[t]{0.235\linewidth}
% \centering    %子图居中
% \includegraphics[scale=0.18]{pic/param_1_pre.pdf}%以pic.jpg的0.5倍大小输出
% \end{minipage}
% }
% % \subfigure[F1-score]{
% \subfloat[F1-score]{
% \begin{minipage}[t]{0.235\linewidth}
% \centering    %子图居中
% \includegraphics[scale=0.18]{pic/param_1_f1.pdf}%以pic.jpg的0.5倍大小输出
% \end{minipage}
% }
% \vspace{+0.5cm}
% \caption{Sensitivity analysis of TTAGN with different sequence lengths}    %大图名称
% \label{fig:param_1}    %图片引用标记
% \end{figure*}

\begin{figure*}[htbp]
\centering  %居中
    \begin{subfigure}{0.235\textwidth}
        \includegraphics[scale=0.18]{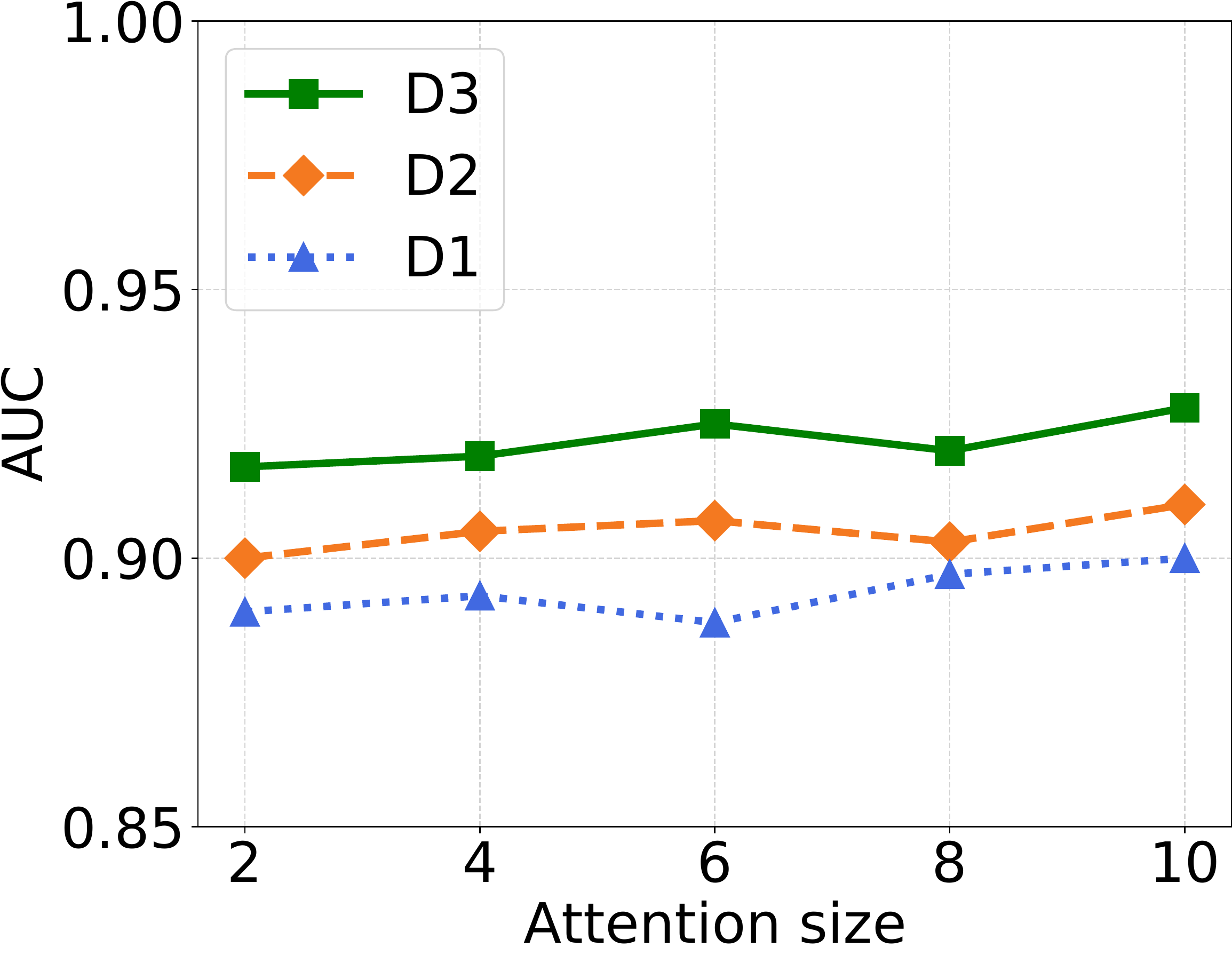}
        \caption{AUC}
    \end{subfigure}
    % \hskip2em
    % \hskip2em
    \begin{subfigure}{0.235\textwidth}
        \includegraphics[scale=0.18]{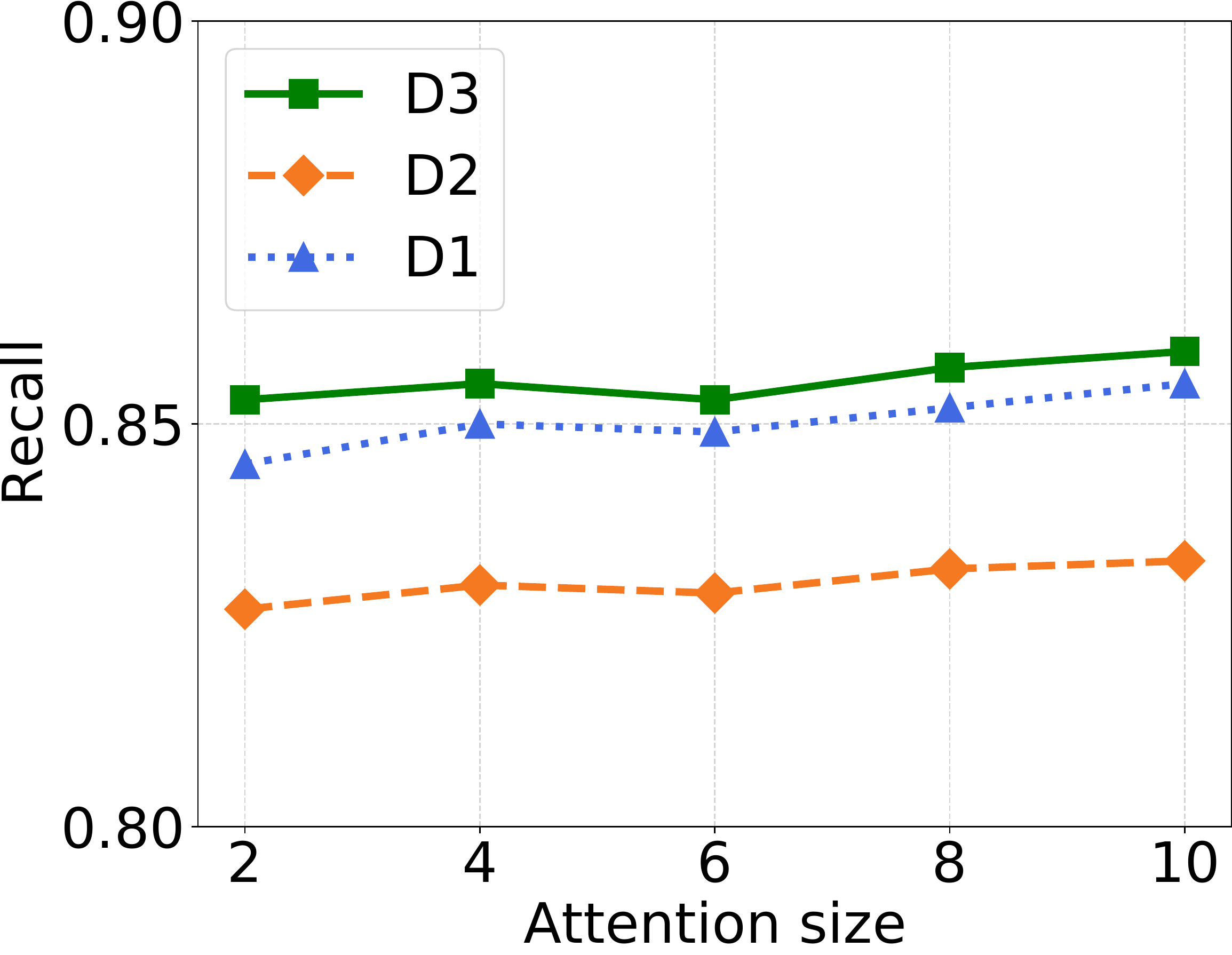}
        \caption{Recall}
    \end{subfigure}
    \begin{subfigure}{0.235\textwidth}
        \includegraphics[scale=0.18]{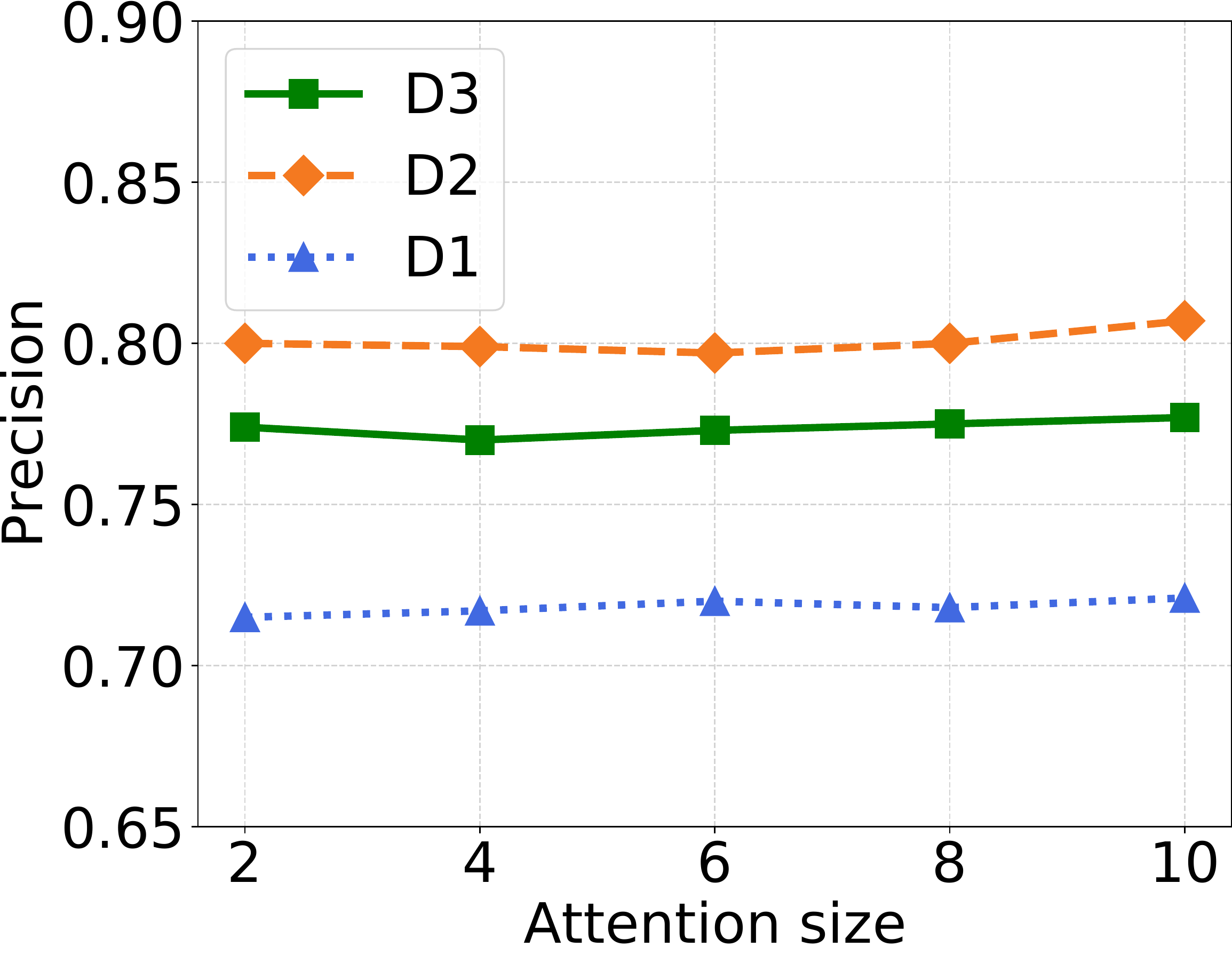}
        \caption{Precision}
    \end{subfigure}
    \begin{subfigure}{0.235\textwidth}
        \includegraphics[scale=0.18]{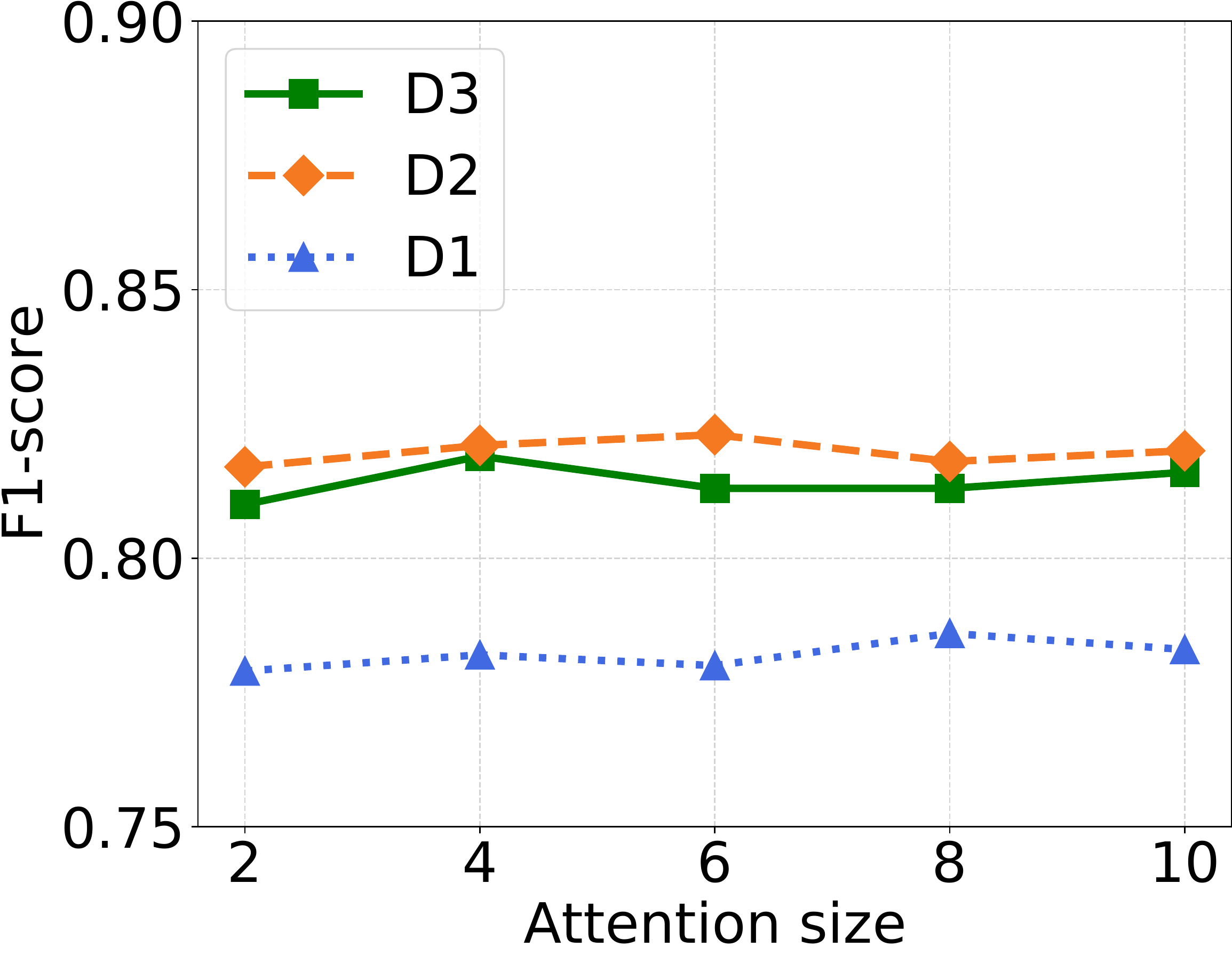}
        \caption{F1-score}
    \end{subfigure}
\vspace{+0.5cm}
\caption{Model robustness study of TTAGN with different attention sizes}    %大图名称
\label{fig:param_2}    %图片引用标记
\end{figure*}

% \begin{figure*}[htbp]
% \centering  %居中
% % \subfigure[AUC]{   %第一张子图
% \subfloat[AUC]{   %第一张子图
% \begin{minipage}[t]{0.235\textwidth}
% \centering    %子图居中
% \includegraphics[scale=0.18]{pic/param_2_auc.pdf}  %以pic.jpg的0.5倍大小输出
% \end{minipage}
% }
% % \subfigure[Recall]{
% \subfloat[Recall]{
% \begin{minipage}[t]{0.235\linewidth}
% \centering    %子图居中
% \includegraphics[scale=0.18]{pic/param_2_recall.pdf}%以pic.jpg的0.5倍大小输出
% \end{minipage}
% }
% % \subfigure[Precision]{ %第二张子图
% \subfloat[Precision]{ %第二张子图
% \begin{minipage}[t]{0.235\linewidth}
% \centering    %子图居中
% \includegraphics[scale=0.18]{pic/param_2_pre.pdf}%以pic.jpg的0.5倍大小输出
% \end{minipage}
% }
% % \subfigure[F1-score]{
% \subfloat[F1-score]{
% \begin{minipage}[t]{0.235\linewidth}
% \centering    %子图居中
% \includegraphics[scale=0.18]{pic/param_2_f1.pdf}%以pic.jpg的0.5倍大小输出
% \end{minipage}
% }
% \vspace{+0.5cm}
% \caption{Model robustness study of TTAGN with different attention sizes}    %大图名称
% \label{fig:param_2}    %图片引用标记
% \end{figure*}

As shown in Figure \ref{fig:ablation}, the corresponding observation results have the following aspects: 

(1) Compared to TTAGN, the performance of TTAGN/t drastically degrades, which are 7\%, 8.5\%, and 9.8\% lower than TTAGN's AUC on $D_{1}$, $D_{2}$, and $D_{3}$ datasets, respectively. The main reason is the sequences model LSTM can fully extract the temporal pattern of transaction interaction between nodes, learn expressive edge representations. This result indicates that learning temporal edges representation of each edge in the transaction graph is essential for the phishing scams detection task, and also proves the importance of edges with transaction information in the transaction graph.
% The effect of TTAGN/t is reduced the most which is TTAGN without temporal edge representation module , and the AUC is lower than TTAGN/e and TTAGN/s by 3.4\%, 6.3\% respectively. 
% and also prove the importance of edges with transaction information in the transaction graph

(2) After removing the edge2node module, TTAGN/e is 6.4\% lower than the full model on the $D_{3}$ dataset. The main function of edge2node is to aggregate around edges representations into nodes. If the learned edge representations are directly spliced with statistical features as classification features, the effect is far inferior to aggregation on nodes. The result proves that the aggregation of edges representations can more comprehensively capture the features of the nodes and strengthen the nodes' representation ability. The edge2node module and the temporal edges representation module complement each other and are indispensable. 
% information biasedly on the nodes

(3) Among the three modules, the structural enhancement module contributes the least. The AUC of TTAGN/s on the $D_{3}$ dataset is 3.1\% lower than TTAGN. It seems that this module is not as significant as the temporal edges representation and edge2node effects, but it also effectively extracts the information of the topological environment. These obtained structural information further enriches the representations of the nodes. 

(4) The performance of the complete model TTAGN on the three datasets is better than other ablation models. This proves that each module could provide effective improvement to finally lead to the significantly high AUC of TTAGN. At the same time, as the graph scale becomes larger, the gap between the models is further widened, proving that TTAGN is more effective on large-scale transaction graphs.

\subsection{Sensitivity Analysis (RQ3)}
To answer RQ3, we further evaluate the performance of TTAGN with respect to the transaction sequence length and edge2node attention size. 

Figure \ref{fig:param_1} presents four metrics scores of TTAGN on three datasets when varying the fixed value of transaction sequence length. The variable-length is also used as a parameter on the far right of the axis. Specifically, we can clearly find that (1) as the fixed value of transaction sequence length increases, combining the four evaluation metrics, the model has achieved enhanced performance on all datasets; (2) Sometimes shorter sequences perform better than the longer sequences, which may be caused by information redundancy; (3) When using the shortest transaction sequence training on the three datasets, there is a large gap compared with the longer transaction sequences; (4) Variable-length performs better than all fixed-length parameters. These phenomena reflect the importance of temporal transaction information and also show that this parameter is unstable, which increasing brings both effective information and information redundancy. Therefore, our method TTAGN proposes the input variable-length transaction sequence, which ensures the effectiveness of the model and also enhances the robustness of the model. 

% \begin{figure}[htbp]
% \centering
% % \includegraphics[width = 1\linewidth]{free-traffic_detection_final.pdf}
% \includegraphics[width = 0.8\linewidth]{param_2_auc.pdf}
% \caption{}
% \label{free-traffic_detection}
% \end{figure}

As for edge2node' attention size, by providing different attention sizes during training, the model sensitivity results are presented in Figure \ref{fig:param_2}. 
% We can observe that, under each attention size setting, TTAGN always achieves the best performance in terms of four evaluation metrics on all datasets. 
We can observe that, under each attention size setting (1) TTAGN always achieves similar performance in terms of four evaluation metrics on all datasets, which is still the best performance compared with other methods in Table \ref{table:compared methods}; 
% The F1 scores of TTAGN and CARE-GNN are comparable and surpass that of GraphConsis by a large margin. 
(2) TTAGN can still achieve relatively good performance when training with a small attention size (e.g., $h$ = 2), which demonstrates the strong capability of its infrastructure. For example, on $D_{3}$ dataset, the Recall barely drops 0.06 if we change the attention size from $h$ = 10 to $h$ = 2. Therefore, we conclude that TTAGN is robust to the edge2node' attention size and consistently outperforms other compared methods.

%% file: conclusion.tex
\section{Conclusion}
\label{sec:conclusion}
In this work, we propose a Temporal Transaction Aggregation Graph Network (TTAGN) to enhance the performance of phishing scams detection on Ethereum. TTAGN fully models and captures the temporal relationship of historical transaction records between nodes, which helps effectively extract edge representations of the Ethereum transaction network. Then, TTAGN aggregates the obtained effective edge representations to fuse topological interactive relationships into nodes, generates trading features which enrich nodes' characteristics and realize their strong representation ability. Finally, combining the three types of features, we improve the performance of Ethereum phishing scams detection. Extensive experiments indicate that TTAGN’s performance and practicality outperform state-of-the-art algorithms by significant margins. We hope that our work demonstrates the serious threat of phishing scams on Ethereum and calls for effective countermeasures deployed by the blockchain community.

% the edge representations around the node are aggregated to fuse topological interactive relationships into its representation, also named as trading features, which enriches the characteristics of the nodes.
% enriching the characteristics of the nodes and 
% fully models and mines temporal information of Ethereum transaction records between transaction nodes to generate the edge representations
% In the first module \emph{Temporal Edge Representation}, we fully mine temporal transaction information from the constructed Ethereum transaction graph, and effectively learn the embedding of transaction edges. In the second module \emph{Edge2node}, we biased aggregate the obtained effective edge embeddings to the nodes, thereby enriching the characteristics of the nodes and realizing the strong representation ability of the nodes. In the third module \emph{Structural Enhancement}, we further learn the structural features of the node and integrate it with the previously learned temporal transaction representation.

%%% Local Variables:
%%% mode: latex
%%% TeX-master: "main"
%%% End:

%% file: acknowledgments.tex
\begin{acks}
This work is supported by The National Key Research and Development Program of China No.2020YFB1006100 and the Strategic Priority Research Program of Chinese Academy of Sciences, Grant No.XDC02040400. We would like to thank the anonymous reviewers for their valuable comments.
\end{acks}

%% file: main.bbl
%%% -*-BibTeX-*-
%%% Do NOT edit. File created by BibTeX with style
%%% ACM-Reference-Format-Journals [18-Jan-2012].

\begin{thebibliography}{31}

%%% ====================================================================
%%% NOTE TO THE USER: you can override these defaults by providing
%%% customized versions of any of these macros before the \bibliography
%%% command.  Each of them MUST provide its own final punctuation,
%%% except for \shownote{}, \showDOI{}, and \showURL{}.  The latter two
%%% do not use final punctuation, in order to avoid confusing it with
%%% the Web address.
%%%
%%% To suppress output of a particular field, define its macro to expand
%%% to an empty string, or better, \unskip, like this:
%%%
%%% \newcommand{\showDOI}[1]{\unskip}   % LaTeX syntax
%%%
%%% \def \showDOI #1{\unskip}           % plain TeX syntax
%%%
%%% ====================================================================

\ifx \showCODEN    \undefined \def \showCODEN     #1{\unskip}     \fi
\ifx \showDOI      \undefined \def \showDOI       #1{#1}\fi
\ifx \showISBNx    \undefined \def \showISBNx     #1{\unskip}     \fi
\ifx \showISBNxiii \undefined \def \showISBNxiii  #1{\unskip}     \fi
\ifx \showISSN     \undefined \def \showISSN      #1{\unskip}     \fi
\ifx \showLCCN     \undefined \def \showLCCN      #1{\unskip}     \fi
\ifx \shownote     \undefined \def \shownote      #1{#1}          \fi
\ifx \showarticletitle \undefined \def \showarticletitle #1{#1}   \fi
\ifx \showURL      \undefined \def \showURL       {\relax}        \fi
% The following commands are used for tagged output and should be
% invisible to TeX
\providecommand\bibfield[2]{#2}
\providecommand\bibinfo[2]{#2}
\providecommand\natexlab[1]{#1}
\providecommand\showeprint[2][]{arXiv:#2}

\bibitem[\protect\citeauthoryear{Ahmed, Shervashidze, Narayanamurthy,
  Josifovski, and Smola}{Ahmed et~al\mbox{.}}{2013}]%
        {ahmed2013distributed}
\bibfield{author}{\bibinfo{person}{Amr Ahmed}, \bibinfo{person}{Nino
  Shervashidze}, \bibinfo{person}{Shravan Narayanamurthy},
  \bibinfo{person}{Vanja Josifovski}, {and} \bibinfo{person}{Alexander~J
  Smola}.} \bibinfo{year}{2013}\natexlab{}.
\newblock \showarticletitle{Distributed large-scale natural graph
  factorization}. In \bibinfo{booktitle}{\emph{Proceedings of the 22nd
  international conference on World Wide Web}}. \bibinfo{pages}{37--48}.
\newblock


\bibitem[\protect\citeauthoryear{Alqassem, Rahwan, and Svetinovic}{Alqassem
  et~al\mbox{.}}{2018}]%
        {alqassem2018anti}
\bibfield{author}{\bibinfo{person}{Israa Alqassem}, \bibinfo{person}{Iyad
  Rahwan}, {and} \bibinfo{person}{Davor Svetinovic}.}
  \bibinfo{year}{2018}\natexlab{}.
\newblock \showarticletitle{The anti-social system properties: Bitcoin network
  data analysis}.
\newblock \bibinfo{journal}{\emph{IEEE Transactions on Systems, Man, and
  Cybernetics: Systems}} \bibinfo{volume}{50}, \bibinfo{number}{1}
  (\bibinfo{year}{2018}), \bibinfo{pages}{21--31}.
\newblock


\bibitem[\protect\citeauthoryear{Anita and Vijayalakshmi}{Anita and
  Vijayalakshmi}{2019}]%
        {survey2019blockchain}
\bibfield{author}{\bibinfo{person}{N Anita} {and} \bibinfo{person}{M
  Vijayalakshmi}.} \bibinfo{year}{2019}\natexlab{}.
\newblock \showarticletitle{Blockchain security attack: a brief survey}. In
  \bibinfo{booktitle}{\emph{2019 10th International Conference on Computing,
  Communication and Networking Technologies (ICCCNT)}}. IEEE,
  \bibinfo{pages}{1--6}.
\newblock


\bibitem[\protect\citeauthoryear{Belkin and Niyogi}{Belkin and Niyogi}{2001}]%
        {laplacian}
\bibfield{author}{\bibinfo{person}{Mikhail Belkin} {and}
  \bibinfo{person}{Partha Niyogi}.} \bibinfo{year}{2001}\natexlab{}.
\newblock \showarticletitle{Laplacian eigenmaps and spectral techniques for
  embedding and clustering.}. In \bibinfo{booktitle}{\emph{Nips}},
  Vol.~\bibinfo{volume}{14}. \bibinfo{pages}{585--591}.
\newblock


\bibitem[\protect\citeauthoryear{Cao, Lu, and Xu}{Cao et~al\mbox{.}}{2015}]%
        {cao2015grarep}
\bibfield{author}{\bibinfo{person}{Shaosheng Cao}, \bibinfo{person}{Wei Lu},
  {and} \bibinfo{person}{Qiongkai Xu}.} \bibinfo{year}{2015}\natexlab{}.
\newblock \showarticletitle{Grarep: Learning graph representations with global
  structural information}. In \bibinfo{booktitle}{\emph{Proceedings of the 24th
  ACM international on conference on information and knowledge management}}.
  \bibinfo{pages}{891--900}.
\newblock


\bibitem[\protect\citeauthoryear{chainalysis}{chainalysis}{[n.\,d.]}]%
        {chainalysis}
\bibfield{author}{\bibinfo{person}{chainalysis}.}
  \bibinfo{year}{[n.\,d.]}\natexlab{}.
\newblock \bibinfo{title}{2021-Crypto-Crime-Report}.
\newblock
  \bibinfo{howpublished}{\url{https://go.chainalysis.com/2021-Crypto-Crime-Report.html}}.
\newblock
\newblock
\shownote{Accessed February, 2021}.


\bibitem[\protect\citeauthoryear{Chen, Pendleton, Njilla, and Xu}{Chen
  et~al\mbox{.}}{2020b}]%
        {2020survey}
\bibfield{author}{\bibinfo{person}{Huashan Chen}, \bibinfo{person}{Marcus
  Pendleton}, \bibinfo{person}{Laurent Njilla}, {and} \bibinfo{person}{Shouhuai
  Xu}.} \bibinfo{year}{2020}\natexlab{b}.
\newblock \showarticletitle{A survey on ethereum systems security:
  Vulnerabilities, attacks, and defenses}.
\newblock \bibinfo{journal}{\emph{ACM Computing Surveys (CSUR)}}
  \bibinfo{volume}{53}, \bibinfo{number}{3} (\bibinfo{year}{2020}),
  \bibinfo{pages}{1--43}.
\newblock


\bibitem[\protect\citeauthoryear{Chen, Peng, Liu, Li, Xie, and Zheng}{Chen
  et~al\mbox{.}}{2020c}]%
        {chen2020phishinggcn}
\bibfield{author}{\bibinfo{person}{Liang Chen}, \bibinfo{person}{Jiaying Peng},
  \bibinfo{person}{Yang Liu}, \bibinfo{person}{Jintang Li},
  \bibinfo{person}{Fenfang Xie}, {and} \bibinfo{person}{Zibin Zheng}.}
  \bibinfo{year}{2020}\natexlab{c}.
\newblock \showarticletitle{Phishing scams detection in ethereum transaction
  network}.
\newblock \bibinfo{journal}{\emph{ACM Transactions on Internet Technology
  (TOIT)}} \bibinfo{volume}{21}, \bibinfo{number}{1} (\bibinfo{year}{2020}),
  \bibinfo{pages}{1--16}.
\newblock


\bibitem[\protect\citeauthoryear{Chen, Guo, Chen, Zheng, and Lu}{Chen
  et~al\mbox{.}}{2020a}]%
        {chen2020phishing}
\bibfield{author}{\bibinfo{person}{Weili Chen}, \bibinfo{person}{Xiongfeng
  Guo}, \bibinfo{person}{Zhiguang Chen}, \bibinfo{person}{Zibin Zheng}, {and}
  \bibinfo{person}{Yutong Lu}.} \bibinfo{year}{2020}\natexlab{a}.
\newblock \showarticletitle{Phishing Scam Detection on Ethereum: Towards
  Financial Security for Blockchain Ecosystem.}. In
  \bibinfo{booktitle}{\emph{IJCAI}}. \bibinfo{pages}{4506--4512}.
\newblock


\bibitem[\protect\citeauthoryear{Goyal and Ferrara}{Goyal and Ferrara}{2018}]%
        {survey2018graph}
\bibfield{author}{\bibinfo{person}{Palash Goyal} {and} \bibinfo{person}{Emilio
  Ferrara}.} \bibinfo{year}{2018}\natexlab{}.
\newblock \showarticletitle{Graph embedding techniques, applications, and
  performance: A survey}.
\newblock \bibinfo{journal}{\emph{Knowledge-Based Systems}}
  \bibinfo{volume}{151} (\bibinfo{year}{2018}), \bibinfo{pages}{78--94}.
\newblock


\bibitem[\protect\citeauthoryear{Grover and Leskovec}{Grover and
  Leskovec}{2016}]%
        {node2vec}
\bibfield{author}{\bibinfo{person}{Aditya Grover} {and} \bibinfo{person}{Jure
  Leskovec}.} \bibinfo{year}{2016}\natexlab{}.
\newblock \showarticletitle{node2vec: Scalable feature learning for networks}.
  In \bibinfo{booktitle}{\emph{Proceedings of the 22nd ACM SIGKDD international
  conference on Knowledge discovery and data mining}}.
  \bibinfo{pages}{855--864}.
\newblock


\bibitem[\protect\citeauthoryear{Hamilton, Ying, and Leskovec}{Hamilton
  et~al\mbox{.}}{2017}]%
        {hamilton2017graphsage}
\bibfield{author}{\bibinfo{person}{William~L Hamilton}, \bibinfo{person}{Rex
  Ying}, {and} \bibinfo{person}{Jure Leskovec}.}
  \bibinfo{year}{2017}\natexlab{}.
\newblock \showarticletitle{Inductive representation learning on large graphs}.
  In \bibinfo{booktitle}{\emph{Proceedings of the 31st International Conference
  on Neural Information Processing Systems}}. \bibinfo{pages}{1025--1035}.
\newblock


\bibitem[\protect\citeauthoryear{Haruta, Asahina, and Sasase}{Haruta
  et~al\mbox{.}}{2017}]%
        {haruta2017ccs}
\bibfield{author}{\bibinfo{person}{Shuichiro Haruta}, \bibinfo{person}{Hiromu
  Asahina}, {and} \bibinfo{person}{Iwao Sasase}.}
  \bibinfo{year}{2017}\natexlab{}.
\newblock \showarticletitle{Visual similarity-based phishing detection scheme
  using image and CSS with target website finder}. In
  \bibinfo{booktitle}{\emph{GLOBECOM 2017-2017 IEEE Global Communications
  Conference}}. IEEE, \bibinfo{pages}{1--6}.
\newblock


\bibitem[\protect\citeauthoryear{Hochreiter and Schmidhuber}{Hochreiter and
  Schmidhuber}{1997}]%
        {LSTM}
\bibfield{author}{\bibinfo{person}{Sepp Hochreiter} {and}
  \bibinfo{person}{J{\"u}rgen Schmidhuber}.} \bibinfo{year}{1997}\natexlab{}.
\newblock \showarticletitle{Long short-term memory}.
\newblock \bibinfo{journal}{\emph{Neural computation}} \bibinfo{volume}{9},
  \bibinfo{number}{8} (\bibinfo{year}{1997}), \bibinfo{pages}{1735--1780}.
\newblock


\bibitem[\protect\citeauthoryear{Holub and O'Connor}{Holub and
  O'Connor}{2018}]%
        {hotbed}
\bibfield{author}{\bibinfo{person}{Artsiom Holub} {and}
  \bibinfo{person}{Jeremiah O'Connor}.} \bibinfo{year}{2018}\natexlab{}.
\newblock \showarticletitle{COINHOARDER: Tracking a ukrainian bitcoin phishing
  ring DNS style}. In \bibinfo{booktitle}{\emph{2018 APWG Symposium on
  Electronic Crime Research (eCrime)}}. IEEE, \bibinfo{pages}{1--5}.
\newblock


\bibitem[\protect\citeauthoryear{Ke, Meng, Finley, Wang, Chen, Ma, Ye, and
  Liu}{Ke et~al\mbox{.}}{2017}]%
        {lightgbm}
\bibfield{author}{\bibinfo{person}{Guolin Ke}, \bibinfo{person}{Qi Meng},
  \bibinfo{person}{Thomas Finley}, \bibinfo{person}{Taifeng Wang},
  \bibinfo{person}{Wei Chen}, \bibinfo{person}{Weidong Ma},
  \bibinfo{person}{Qiwei Ye}, {and} \bibinfo{person}{Tie-Yan Liu}.}
  \bibinfo{year}{2017}\natexlab{}.
\newblock \showarticletitle{Lightgbm: A highly efficient gradient boosting
  decision tree}.
\newblock \bibinfo{journal}{\emph{Advances in neural information processing
  systems}}  \bibinfo{volume}{30} (\bibinfo{year}{2017}),
  \bibinfo{pages}{3146--3154}.
\newblock


\bibitem[\protect\citeauthoryear{Kipf and Welling}{Kipf and Welling}{2016a}]%
        {GCN}
\bibfield{author}{\bibinfo{person}{Thomas~N Kipf} {and} \bibinfo{person}{Max
  Welling}.} \bibinfo{year}{2016}\natexlab{a}.
\newblock \showarticletitle{Semi-supervised classification with graph
  convolutional networks}.
\newblock \bibinfo{journal}{\emph{arXiv preprint arXiv:1609.02907}}
  (\bibinfo{year}{2016}).
\newblock


\bibitem[\protect\citeauthoryear{Kipf and Welling}{Kipf and Welling}{2016b}]%
        {kipf2016vgae}
\bibfield{author}{\bibinfo{person}{Thomas~N Kipf} {and} \bibinfo{person}{Max
  Welling}.} \bibinfo{year}{2016}\natexlab{b}.
\newblock \showarticletitle{Variational graph auto-encoders}.
\newblock \bibinfo{journal}{\emph{arXiv preprint arXiv:1611.07308}}
  (\bibinfo{year}{2016}).
\newblock


\bibitem[\protect\citeauthoryear{Lin, Wu, Yuan, and Zheng}{Lin
  et~al\mbox{.}}{2020}]%
        {lin2020modeling}
\bibfield{author}{\bibinfo{person}{Dan Lin}, \bibinfo{person}{Jiajing Wu},
  \bibinfo{person}{Qi Yuan}, {and} \bibinfo{person}{Zibin Zheng}.}
  \bibinfo{year}{2020}\natexlab{}.
\newblock \showarticletitle{Modeling and understanding ethereum transaction
  records via a complex network approach}.
\newblock \bibinfo{journal}{\emph{IEEE Transactions on Circuits and Systems II:
  Express Briefs}} \bibinfo{volume}{67}, \bibinfo{number}{11}
  (\bibinfo{year}{2020}), \bibinfo{pages}{2737--2741}.
\newblock


\bibitem[\protect\citeauthoryear{Ou, Cui, Pei, Zhang, and Zhu}{Ou
  et~al\mbox{.}}{2016}]%
        {HOPE}
\bibfield{author}{\bibinfo{person}{Mingdong Ou}, \bibinfo{person}{Peng Cui},
  \bibinfo{person}{Jian Pei}, \bibinfo{person}{Ziwei Zhang}, {and}
  \bibinfo{person}{Wenwu Zhu}.} \bibinfo{year}{2016}\natexlab{}.
\newblock \showarticletitle{Asymmetric transitivity preserving graph
  embedding}. In \bibinfo{booktitle}{\emph{Proceedings of the 22nd ACM SIGKDD
  international conference on Knowledge discovery and data mining}}.
  \bibinfo{pages}{1105--1114}.
\newblock


\bibitem[\protect\citeauthoryear{Perozzi, Al-Rfou, and Skiena}{Perozzi
  et~al\mbox{.}}{2014}]%
        {deepwalk}
\bibfield{author}{\bibinfo{person}{Bryan Perozzi}, \bibinfo{person}{Rami
  Al-Rfou}, {and} \bibinfo{person}{Steven Skiena}.}
  \bibinfo{year}{2014}\natexlab{}.
\newblock \showarticletitle{Deepwalk: Online learning of social
  representations}. In \bibinfo{booktitle}{\emph{Proceedings of the 20th ACM
  SIGKDD international conference on Knowledge discovery and data mining}}.
  \bibinfo{pages}{701--710}.
\newblock


\bibitem[\protect\citeauthoryear{Roweis and Saul}{Roweis and Saul}{2000}]%
        {LLE}
\bibfield{author}{\bibinfo{person}{Sam~T Roweis} {and}
  \bibinfo{person}{Lawrence~K Saul}.} \bibinfo{year}{2000}\natexlab{}.
\newblock \showarticletitle{Nonlinear dimensionality reduction by locally
  linear embedding}.
\newblock \bibinfo{journal}{\emph{science}} \bibinfo{volume}{290},
  \bibinfo{number}{5500} (\bibinfo{year}{2000}), \bibinfo{pages}{2323--2326}.
\newblock


\bibitem[\protect\citeauthoryear{Sahingoz, Buber, Demir, and Diri}{Sahingoz
  et~al\mbox{.}}{2019}]%
        {sahingoz2019url}
\bibfield{author}{\bibinfo{person}{Ozgur~Koray Sahingoz},
  \bibinfo{person}{Ebubekir Buber}, \bibinfo{person}{Onder Demir}, {and}
  \bibinfo{person}{Banu Diri}.} \bibinfo{year}{2019}\natexlab{}.
\newblock \showarticletitle{Machine learning based phishing detection from
  URLs}.
\newblock \bibinfo{journal}{\emph{Expert Systems with Applications}}
  \bibinfo{volume}{117} (\bibinfo{year}{2019}), \bibinfo{pages}{345--357}.
\newblock


\bibitem[\protect\citeauthoryear{Tang, Qu, Wang, Zhang, Yan, and Mei}{Tang
  et~al\mbox{.}}{2015}]%
        {line}
\bibfield{author}{\bibinfo{person}{Jian Tang}, \bibinfo{person}{Meng Qu},
  \bibinfo{person}{Mingzhe Wang}, \bibinfo{person}{Ming Zhang},
  \bibinfo{person}{Jun Yan}, {and} \bibinfo{person}{Qiaozhu Mei}.}
  \bibinfo{year}{2015}\natexlab{}.
\newblock \showarticletitle{Line: Large-scale information network embedding}.
  In \bibinfo{booktitle}{\emph{Proceedings of the 24th international conference
  on world wide web}}. \bibinfo{pages}{1067--1077}.
\newblock


\bibitem[\protect\citeauthoryear{Vaswani, Shazeer, Parmar, Uszkoreit, Jones,
  Gomez, Kaiser, and Polosukhin}{Vaswani et~al\mbox{.}}{2017}]%
        {vaswani2017attention}
\bibfield{author}{\bibinfo{person}{Ashish Vaswani}, \bibinfo{person}{Noam
  Shazeer}, \bibinfo{person}{Niki Parmar}, \bibinfo{person}{Jakob Uszkoreit},
  \bibinfo{person}{Llion Jones}, \bibinfo{person}{Aidan~N Gomez},
  \bibinfo{person}{{\L}ukasz Kaiser}, {and} \bibinfo{person}{Illia
  Polosukhin}.} \bibinfo{year}{2017}\natexlab{}.
\newblock \showarticletitle{Attention is all you need}. In
  \bibinfo{booktitle}{\emph{Advances in neural information processing
  systems}}. \bibinfo{pages}{5998--6008}.
\newblock


\bibitem[\protect\citeauthoryear{Veli{\v{c}}kovi{\'c}, Cucurull, Casanova,
  Romero, Lio, and Bengio}{Veli{\v{c}}kovi{\'c} et~al\mbox{.}}{2017}]%
        {velivckovic2017gat}
\bibfield{author}{\bibinfo{person}{Petar Veli{\v{c}}kovi{\'c}},
  \bibinfo{person}{Guillem Cucurull}, \bibinfo{person}{Arantxa Casanova},
  \bibinfo{person}{Adriana Romero}, \bibinfo{person}{Pietro Lio}, {and}
  \bibinfo{person}{Yoshua Bengio}.} \bibinfo{year}{2017}\natexlab{}.
\newblock \showarticletitle{Graph attention networks}.
\newblock \bibinfo{journal}{\emph{arXiv preprint arXiv:1710.10903}}
  (\bibinfo{year}{2017}).
\newblock


\bibitem[\protect\citeauthoryear{Wang, Cui, and Zhu}{Wang
  et~al\mbox{.}}{2016}]%
        {wang2016sdne}
\bibfield{author}{\bibinfo{person}{Daixin Wang}, \bibinfo{person}{Peng Cui},
  {and} \bibinfo{person}{Wenwu Zhu}.} \bibinfo{year}{2016}\natexlab{}.
\newblock \showarticletitle{Structural deep network embedding}. In
  \bibinfo{booktitle}{\emph{Proceedings of the 22nd ACM SIGKDD international
  conference on Knowledge discovery and data mining}}.
  \bibinfo{pages}{1225--1234}.
\newblock


\bibitem[\protect\citeauthoryear{Wang, Jin, Dai, Choo, and Zou}{Wang
  et~al\mbox{.}}{2021}]%
        {wang2021ethereumsurvey}
\bibfield{author}{\bibinfo{person}{Zeli Wang}, \bibinfo{person}{Hai Jin},
  \bibinfo{person}{Weiqi Dai}, \bibinfo{person}{Kim-Kwang~Raymond Choo}, {and}
  \bibinfo{person}{Deqing Zou}.} \bibinfo{year}{2021}\natexlab{}.
\newblock \showarticletitle{Ethereum smart contract security research: survey
  and future research opportunities}.
\newblock \bibinfo{journal}{\emph{Frontiers of Computer Science}}
  \bibinfo{volume}{15}, \bibinfo{number}{2} (\bibinfo{year}{2021}),
  \bibinfo{pages}{1--18}.
\newblock


\bibitem[\protect\citeauthoryear{Wood et~al\mbox{.}}{Wood
  et~al\mbox{.}}{2014}]%
        {wood2014ethereum}
\bibfield{author}{\bibinfo{person}{Gavin Wood} {et~al\mbox{.}}}
  \bibinfo{year}{2014}\natexlab{}.
\newblock \showarticletitle{Ethereum: A secure decentralised generalised
  transaction ledger}.
\newblock \bibinfo{journal}{\emph{Ethereum project yellow paper}}
  \bibinfo{volume}{151}, \bibinfo{number}{2014} (\bibinfo{year}{2014}),
  \bibinfo{pages}{1--32}.
\newblock


\bibitem[\protect\citeauthoryear{Wu, Yuan, Lin, You, Chen, Chen, and Zheng}{Wu
  et~al\mbox{.}}{2020}]%
        {wu2020phishers}
\bibfield{author}{\bibinfo{person}{Jiajing Wu}, \bibinfo{person}{Qi Yuan},
  \bibinfo{person}{Dan Lin}, \bibinfo{person}{Wei You}, \bibinfo{person}{Weili
  Chen}, \bibinfo{person}{Chuan Chen}, {and} \bibinfo{person}{Zibin Zheng}.}
  \bibinfo{year}{2020}\natexlab{}.
\newblock \showarticletitle{Who are the phishers? phishing scam detection on
  ethereum via network embedding}.
\newblock \bibinfo{journal}{\emph{IEEE Transactions on Systems, Man, and
  Cybernetics: Systems}} (\bibinfo{year}{2020}).
\newblock


\bibitem[\protect\citeauthoryear{Yuan, Huang, Zhang, Wu, Zhang, and Zhang}{Yuan
  et~al\mbox{.}}{2020}]%
        {yuan2020detecting}
\bibfield{author}{\bibinfo{person}{Qi Yuan}, \bibinfo{person}{Baoying Huang},
  \bibinfo{person}{Jie Zhang}, \bibinfo{person}{Jiajing Wu},
  \bibinfo{person}{Haonan Zhang}, {and} \bibinfo{person}{Xi Zhang}.}
  \bibinfo{year}{2020}\natexlab{}.
\newblock \showarticletitle{Detecting phishing scams on ethereum based on
  transaction records}. In \bibinfo{booktitle}{\emph{2020 IEEE International
  Symposium on Circuits and Systems (ISCAS)}}. IEEE, \bibinfo{pages}{1--5}.
\newblock


\end{thebibliography}
